%% file: SMP-13-002_temp.tex
\pdfoutput=1

\documentclass[11pt,twoside,a4paper,cmspaper,final,collab]{cms-tdr}
\def\svnVersion{265421}\def\cmsCernNo{2013-037}\def\cmsCernDate{\today}\def\cmsMessage{Published in Physical Review D as \href{http://dx.doi.org/10.1103/PhysRevD.90.072006}{\doi{10.1103/PhysRevD.90.072006.}}}

\begin{document}\cmsNoteHeader{SMP-13-002}

\hyphenation{had-ron-i-za-tion}
\hyphenation{cal-or-i-me-ter}
\hyphenation{de-vices}
\RCS$Revision: 265381 $
\RCS$HeadURL: svn+ssh://svn.cern.ch/reps/tdr2/papers/SMP-13-002/trunk/SMP-13-002.tex $
\RCS$Id: SMP-13-002.tex 265381 2014-10-28 12:16:06Z mgouzevi $
\newlength\cmsFigWidth
\ifthenelse{\boolean{cms@external}}{\setlength\cmsFigWidth{0.90\textwidth}}{\setlength\cmsFigWidth{\textwidth}}
\ifthenelse{\boolean{cms@external}}{\providecommand{\cmsLeft}{top}}{\providecommand{\cmsLeft}{left}}
\ifthenelse{\boolean{cms@external}}{\providecommand{\cmsRight}{bottom}}{\providecommand{\cmsRight}{right}}
\providecommand{\ownint}[4]{{\int_{#1}^{#2} \! #3 \, \rd#4}}
\providecommand{\E}[1]{{\operatorname{E} \left[ #1 \right]}}
\providecommand*{\pd}[2]{\frac{\partial #1}{\partial #2}}
\providecommand*{\var}[1]{\operatorname{Var}[ #1 ]}
\providecommand{\R}{\ensuremath{\mathcal{R}}\xspace}
\providecommand{\Leff}{\ensuremath{\mathcal{L}_\text{eff}}\xspace}
\providecommand{\Lint}{\ensuremath{\mathcal{L}_\text{int}}\xspace}
\providecommand{\shat}{\ensuremath{\hat{\sigma}}\xspace}
\providecommand{\stilde}{\ensuremath{\tilde{\sigma}}\xspace}
\providecommand{\alps}{\ensuremath{\alpha_S}\xspace}
\providecommand{\alpsmz}{\ensuremath{\alpha_S(M_Z)}\xspace}
\providecommand{\fastjet}{{\textsc{FastJet}}\xspace}
\providecommand{\fastNLO}{{\textsc{fastNLO}}\xspace}
\providecommand{\HERWIGPP}{{\textsc{herwig++}}\xspace}
\providecommand{\MADGRAPHF}{\textsc{MadGraph5}\xspace}
\providecommand{\NLOJETPP}{{\textsc{NLOJet++}}\xspace}
\providecommand{\POWHEGBOX}{{\textsc{powheg box}}\xspace}
\providecommand{\PYTHIAS}{{\textsc{pythia6}}\xspace}
\providecommand{\PYTHIAE}{{\textsc{pythia8}}\xspace}
\providecommand{\RooUnfold}{{\textsc{RooUnfold}}\xspace}
\ifthenelse{\boolean{cms@external}}{\providecommand{\cmsResizeTable}[1]{\resizebox{\columnwidth}{!}{#1}}}{\providecommand{\cmsResizeTable[1]}{\relax #1}}
\cmsNoteHeader{SMP-13-002} % This is over-written in the CMS environment: useful as preprint no. for export versions
\title{Measurement of the ratio of inclusive jet cross sections using
  the \texorpdfstring{anti-\kt}{anti-kt} algorithm with radius parameters \texorpdfstring{$R=0.5$ and
  0.7 in $\Pp\Pp$ collisions at $\sqrt{s} = 7\TeV$}{R = 0.5 and 0.7 in pp collisions at sqrt(s) = 7 TeV}}

\date{\today}

\abstract{Measurements of the inclusive jet cross section with the
  anti-\kt clustering algorithm are presented for two radius
  parameters, $R=0.5$ and 0.7. They are based on data from LHC
  proton-proton collisions at $\sqrt{s}= 7\TeV$ corresponding to an
  integrated luminosity of 5.0\fbinv collected with the CMS
  detector in 2011. The ratio of these two measurements is obtained
  as a function of the rapidity and transverse momentum of the jets.
  Significant discrepancies are found comparing the data to
  leading-order simulations and to fixed-order calculations at
  next-to-leading order, corrected for nonperturbative effects,
  whereas simulations with next-to-leading-order matrix elements
  matched to parton showers describe the data best.}

\hypersetup{%
pdfauthor={CMS Collaboration},%
pdftitle={Measurement of the ratio of inclusive jet cross sections
  using the anti-kt algorithm with radius parameters R = 0.5 and 0.7 in pp collisions at sqrt(s) = 7 TeV},%
pdfsubject={CMS},%
pdfkeywords={CMS, physics, QCD, jet, parton shower}}

\maketitle
\section{Introduction}%
\label{sec-intro}

The inclusive cross section for jets produced with high transverse
momenta in proton-proton collisions is described by quantum
chromodynamics~(QCD) in terms of parton-parton scattering.  The
partonic cross section $\shat_\text{jet}$ is convolved with the
parton distribution functions~(PDFs) of the proton and is computed in
perturbative QCD~(pQCD) as an expansion in powers of the strong
coupling constant, \alps.  In practice, the complexity of the
calculations requires a truncation of the series after a few terms.
Next-to-leading order~(NLO) calculations of inclusive jet and dijet
production were carried out in the early
1990s~\cite{Ellis:1990ek,Ellis:1992en,Giele:1993dj}, and more
recently, progress towards next-to-next-to-leading order (NNLO)
calculations has been reported~\cite{Currie:2013dwa}.

Jet cross sections at the parton level are not well defined unless one
uses a jet algorithm that is safe from collinear and infrared
divergences, \ie, an algorithm that produces a cluster result that
does not change in the presence of soft gluon emissions or collinear
splittings of partons. Analyses conducted with LHC data employ the
anti-\kt jet algorithm~\cite{Cacciari:2008gp}, which is collinear- and
infrared-safe.  At the Tevatron, however, only a subset of analyses
done with the \kt~jet
algorithm~\cite{Catani:1991hj,Brown:1991hx,Catani:1992zp,Ellis:1993tq}
are collinear- and infrared-safe.  Nonetheless, the inclusive jet
measurements with jet size parameters~$R$ on the order of unity
performed by the
CDF~\cite{Affolder:2001fa,Abulencia:2007ez,Aaltonen:2008eq} and
D0~\cite{Abbott:2000ew,Abazov:2001hb,Abazov:2011vi} Collaborations at
1.8 and 1.96\TeV center-of-mass energies are well described by NLO
QCD calculations.
Even though calculations at NLO provide at most three partons in the
final state for jet clustering, measurements with somewhat smaller
anti-\kt jet radii of $R=0.4$ up to 0.7 by the
ATLAS~\cite{Aad:2011fc,Aad:2013lpa},
CMS~\cite{Chatrchyan:2011ab,Chatrchyan:2012gw,Chatrchyan:2012bja}, and
ALICE~\cite{Abelev:2013fn} Collaborations are equally well
characterized for 2.76 and $7\TeV$ center-of-mass energies at the
LHC.

The relative normalization of measured cross sections and theoretical
predictions for different jet radii $R$ exhibits a dependence on~$R$.
This effect has been investigated theoretically in
Refs.~\cite{Dasgupta:2007wa,Cacciari:2008gd}, where it was found that,
in a collinear approximation, the impact of perturbative radiation and
of the nonperturbative effects of hadronization and the underlying
event on jet transverse momenta scales for small $R$ roughly with $\ln
R$, $-1/R$, and $R^2$ respectively. As a consequence, the choice of
the jet radius parameter~$R$ determines which aspects of jet formation
are emphasized.  In order to gain insight into the interplay of these
effects, Ref.~\cite{Dasgupta:2007wa} suggested a study of the relative
difference between inclusive jet cross sections that emerge from two
different jet definitions:
\begin{equation}
  \label{eq:Rdef}
  \left(\frac{\rd\sigma^\text{alt}}{\rd\pt} -
    \frac{\rd\sigma^\text{ref}}{\rd\pt}\right)
  \left/ \left(\frac{\rd\sigma^\text{ref}}{\rd\pt}\right)\right. =
  \R(\text{alt, ref}) - 1.
\end{equation}
Different jet algorithms applied to leading-order (LO) two-parton
final states lead to identical results, provided partons in opposite
hemispheres are not clustered together.  Therefore, the numerator
differs from zero only for three or more partons, and the quantity
defined in Eq.~(\ref{eq:Rdef}) defines a three-jet observable that is
calculable to NLO with terms up to $\alps^4$ with
\NLOJETPP~\cite{Nagy:2001fj,Nagy:2003tz} as demonstrated in
Ref.~\cite{Soyez:2011np}.

The analysis presented here focuses on the study of the jet radius
ratio, $\R(0.5,0.7)$, as a function of the jet \pt and rapidity $y$,
using the anti-\kt jet algorithm with $R = 0.5$ as the alternative and
$R = 0.7$ as the reference jet radius. It is expected that QCD
radiation reduces this ratio below unity and that the effect vanishes
with the increasing collimation of jets at high~\pt.

The LO Monte Carlo (MC) event generators
\PYTHIAS~\cite{Sjostrand:2006za} and \HERWIGPP~\cite{Bahr:2008pv} are
used as a basis for comparison, including parton showers~(PS) and
models for hadronization and the underlying event. As in the previous
publication~\cite{Chatrchyan:2012bja}, they are also used to derive
nonperturbative~(NP) correction factors for the fixed-order
predictions, which will be denoted LO$\otimes$NP and NLO$\otimes$NP as
appropriate. In addition, jet production as predicted with \POWHEG at
NLO~\cite{Alioli:2010xa} and matched to the~PS of \PYTHIAS is compared
to measurements.

A similar study has been performed by the ALICE
Collaboration~\cite{Abelev:2013fn}, and the ZEUS Collaboration at the
HERA collider investigated the jet ratio as defined with two different
jet algorithms~\cite{Abramowicz:2010ke}. Comparisons to predictions
involving \POWHEG have been presented previously by
ATLAS~\cite{Aad:2011fc}.

\section{The CMS detector} \label{sec-detector}

A detailed description of the CMS experiment can be found
elsewhere~\cite{Chatrchyan:2008aa}.  The CMS coordinate system has the origin at
the center of the detector.  The $z$-axis points along the direction
of the counterclockwise beam, with the transverse plane perpendicular
to the beam.  Azimuthal angle is denoted $\phi$, polar angle $\theta$
and
pseudorapidity is defined as $\eta \equiv -\ln(\tan[\theta/2])$.

The central feature of the CMS apparatus is a superconducting
solenoid, of 6\unit{m} internal diameter, providing a field of 3.8\unit{T}. Within
the field volume are a silicon pixel and strip tracker, a crystal
electromagnetic calorimeter (ECAL) and a sampling hadron calorimeter
(HCAL).  The ECAL is made up of lead tungstate crystals, while the
HCAL is made up of layers of plates of brass and plastic
scintillator. These calorimeters provide coverage up to $\abs{\eta}< 3.0$.
An iron and quartz-fiber Cherenkov hadron forward (HF) calorimeter
covers $3.0 < \abs{\eta}< 5.0$.  The muons are measured in the range
$\abs{\eta}< 2.4$, with detection planes made using three technologies:
drift tubes, cathode strip chambers, and resistive-plate chambers.

\section{Jet reconstruction}

The particle-flow (PF) event reconstruction algorithm is meant to
reconstruct and identify each single particle with an optimal
combination of all subdetector information~\cite{CMS-PAS-PFT-09-001}.
The energy of photons is directly obtained from the ECAL measurement,
corrected for zero-suppression effects. The energy of electrons is
determined from a combination of the track momentum at the main
interaction vertex, the corresponding ECAL cluster energy, and the
energy sum of all bremsstrahlung photons attached to the track.  Muons
are identified with the muon system and their energy is obtained from
the corresponding track momentum.  The energy of charged hadrons is
determined from a combination of the track momentum and the
corresponding ECAL and HCAL energy, corrected for zero-suppression
effects, and calibrated for the nonlinear response of the
calorimeters. Finally the energy of neutral hadrons is obtained from
the corresponding calibrated ECAL and HCAL energy.

Jets are reconstructed offline from the PF
objects, clustered by the anti-\kt algorithm with jet radius $R=0.5$
and 0.7 using the \fastjet package~\cite{Cacciari:2011ma}.
The jet momentum is determined as the vectorial sum of all particle
momenta in the jet.
An offset correction is applied to take into account the extra energy
clustered into jets due to additional proton-proton interactions within
the same bunch crossing. Jet energy corrections are derived from the
simulation separately for $R=0.5$ and 0.7 jets, and are confirmed by
in situ measurements with the energy balance of dijet, Z+jet, and
photon+jet events
using the missing $\ET$ projection fraction method, which is
independent of the jet clustering algorithm~\cite{Chatrchyan:2011ds}.  Additional
selection criteria are applied to each event to remove spurious
jet-like features originating from isolated noise patterns in certain
HCAL regions.

The offset correction is particularly important for the jet radius
ratio analysis, because it scales with the jet area, which is on
average twice as large for $R=0.7$ jets than for 0.5 jets, while
most other jet energy
uncertainties cancel out. The offset subtraction is performed with the
hybrid jet area method presented in Ref.~\cite{Chatrchyan:2011ds}. In the original jet
area method~\cite{Cacciari:2007fd} the offset is calculated as a
product of the global energy density $\rho$ and the jet area $A_\text{jet}$, both of which are determined using \fastjet.
In the hybrid method $\rho$ is corrected for:

(1) the experimentally determined $\eta$-dependence of the offset
energy density using minimum bias data, (2) the underlying event
energy density using dijet data, and (3) the difference in offset
energy density inside and outside of the jet cone using simulation.

The average number of pileup interactions in 2011 was between 7.4 and
10.3, depending on the trigger conditions (as discussed in
Sec.~\ref{sec:trigger}). This corresponds to between 5.6 and 7.5 good,
reconstructed vertices, amounting to a pileup vertex reconstruction
and identification efficiency of about 60--65\%.
The global average energy density $\rho$ was between 4.8 and 6.2
\GeV/rad$^2$, averaging to about 0.5\GeV/rad$^2$ per pileup interaction
on top of 1.5\GeV/rad$^2$ for the underlying event, noise, and
out-of-time contributions.
The anti-\kt jet areas are well approximated by $\pi R^2$ and are
about 0.8 and 1.5\unit{rad$^2$} for $R=0.5$ and 0.7, respectively. This
sets the typical offset in
the range of 3.8--4.9\GeV (7.2--9.3\GeV) for $R=0.5$ (0.7).
Most of the pileup offset is due to collisions within the same bunch
crossing, with lesser contributions from neighboring bunch crossings,
\ie out-of-time pileup.
\section{Monte Carlo models and theoretical
  calculations}

Three MC generators are used for simulating events and for theoretical
predictions:
\begin{itemize}
\item \PYTHIA version~6.422~\cite{Sjostrand:2006za} uses LO matrix
  elements to generate the 2$\rightarrow$2 hard process in pQCD and a
  PS model for parton emissions close in phase
  space~\cite{Bengtsson:1986et,Bengtsson:1986hr,Sjostrand:2004ef}. To
  simulate the underlying event several options are
  available~\cite{Sjostrand:1987su,Sjostrand:2004pf,Sjostrand:2004ef}.
  Hadronization is performed with the Lund string
  fragmentation~\cite{Andersson:1983ia,Andersson:1983jt,Sjostrand:1984iu}.
  In this analysis, events are generated with the Z2 tune, where
  parton showers are ordered in \pt. The Z2 tune is identical to the
  Z1 tune described in Ref.~\cite{tunesZ2}, except that Z2 uses the
  CTEQ6L1~\cite{Pumplin:2002vw} parton distribution functions.
\item Similarly, \HERWIGPP is a MC event generator with LO matrix
  elements, which is employed here in the form of version~2.4.2 with
  the default tune of version~2.3~\cite{Bahr:2008pv}. \HERWIGPP
  simulates parton showers using the coherent branching algorithm with
  angular ordering of
  emissions~\cite{Marchesini:1987cf,Gieseke:2003rz}. The underlying
  event is simulated with the eikonal multiple partonic-scattering
  model~\cite{Bahr:2008dy} and hadrons are formed from quarks and
  gluons using cluster fragmentation~\cite{Webber:1983if}.
\item In contrast, the
  \POWHEGBOX~\cite{Nason:2004rx,Frixione:2007vw,Alioli:2010xd} is a
  general computing framework to interface NLO calculations to MC
  event generators. The jet production relevant here is described in
  Ref.~\cite{Alioli:2010xa}. To complete the event generation with
  parton showering, modelling of the underlying event, and
  hadronization, \PYTHIAS was employed in this study, although
  \HERWIGPP can be used as well.
\end{itemize}

All three event generation schemes are compared at particle level to
the jet radius ratio \R. Any dependence of jet production on the jet
radius is generated only through parton showering in \PYTHIAS and
\HERWIGPP, whereas with \POWHEG the hardest additional emission is
provided at the level of the matrix elements.

A fixed-order prediction at LO of the jet radius ratio is obtained
using the \NLOJETPP program version
4.1.3~\cite{Nagy:2001fj,Nagy:2003tz} within the framework of the
\fastNLO package version~2.1~\cite{Britzger:2012bs}. The NLO
calculations are performed using the technique from
Ref.~\cite{Soyez:2011np}. The nonperturbative correction factors are
estimated from \PYTHIAS and \HERWIGPP as in
Ref.~\cite{Chatrchyan:2012bja}.
\section{Measurement of differential inclusive jet cross sections}

The measurement of the jet radius ratio
$\R(0.5,0.7)$ is calculated by forming the ratio of two separate
measurements
of the differential jet cross sections with the anti-\kt clustering
parameters $R=0.5$ and 0.7. These measurements are reported in six
0.5-wide bins of absolute rapidity for $\abs{y} < 3.0$ starting from
$\pt>56\GeV$ for the lowest single jet trigger threshold. The methods
used in this paper closely follow those presented in
Ref.~\cite{Chatrchyan:2012bja} for $R=0.7$, and the results fully agree with the
earlier publication within the overlapping phase space. The results
for $R=0.5$ also agree with the earlier CMS publication~\cite{Chatrchyan:2011ab}
within statistical and systematic
uncertainties. Particular care is taken to ensure that any residual
biases in the $R=0.5$ and 0.7 measurements cancel for the
jet radius ratio, whether coming from the jet energy scale, jet
resolutions, unfolding, trigger, or the integrated luminosity
measurement. The statistical correlations between the two
measurements are properly taken into account, and are propagated to
the final uncertainty estimates for the jet radius ratio~$\R$.

\subsection{Data samples and event selection}\label{sec:trigger}

Events were collected online with a two-tiered trigger system,
consisting of a hardware level-1 and a software high-level
trigger (HLT). The jet algorithm run by the trigger uses the energies
measured in the ECAL, HCAL, and HF calorimeters. The anti-\kt
clustering with radius parameter $R=0.5$ is used as implemented in the
\fastjet package.  The data samples used for this measurement were
collected with single-jet HLT triggers, where in each event at least
one $R=0.5$ jet, measured from calorimetric energies alone, is
required to exceed a minimal \pt as listed in Table~\ref{tab:trigger}.
The triggers with low \pt thresholds have been prescaled to limit the trigger rates, which means that they correspond to a lower integrated luminosity $\mathcal{L}_{\rm int}$, as shown in Table~\ref{tab:trigger}.

The \pt thresholds in the later analysis are substantially higher than
in the HLT to account for differences between jets measured with only
the calorimetric detectors and PF jets.  For each trigger threshold
the efficiency turn-on as a function of \pt for the larger radius
parameter $R=0.7$ is less sharp than for $R=0.5$. This is caused by
potential splits of one $R=0.7$ jet into two $R=0.5$ jets and by
additional smearing from pileup for the larger cone size.
The selection criteria ensure trigger efficiencies above 97\% (98.5\%)
for $R=0.7$ at $\pt=56\GeV$ ($\pt>114\GeV$ as
in Ref.~\cite{Chatrchyan:2012bja}), and above 99.5\% for $R=0.5$ at
$\pt=56\GeV$. The analysis \pt thresholds, which closely follow those
reported in Ref.~\cite{Chatrchyan:2012bja}, are reproduced in
Table~\ref{tab:trigger}.

\begin{table*}[hbtp!]
  \centering
  \topcaption{The trigger and analysis \pt thresholds together with the respective integrated luminosities \Lint.
    \label{tab:trigger}}
  \begin{scotch}{c|cccccc}
    Trigger \pt threshold (\GeVns{})     &     30 &    60 &  110 & 190 & 240 &  300\\\hline
    Minimum \pt for analysis (\GeVns{})  &     56 &    97 &  174 & 300 & 362 &  507\\
    \Lint (\pbinv)                   & 0.0149 & 0.399 & 7.12 & 150 & 513 & 4960\\
  \end{scotch}
\end{table*}

\subsection{Measurement of the cross sections and jet radius ratio}

The jet $\pt$ spectrum is obtained by populating each bin with the
number of jets from the events collected with the associated trigger
as described in the previous section. The yields collected with each
trigger are then scaled according to the respective integrated
luminosity as shown in Table~\ref{tab:trigger}.

The observed inclusive jet yields are transformed into a
double-differential cross section as follows:
\begin{equation}
  \frac{\rd^2\stilde}{\rd\pt \rd{}y} = \frac{1}{\epsilon\cdot\Lint}
  \frac{N_\text{jets}}{\Delta\pt\Delta y},
\end{equation}
where $N_\text{jets}$ is the number of jets in the bin, $\Lint$ is
the integrated luminosity of the data sample from which the events are
taken, $\epsilon$ is the product of the trigger and event selection
efficiencies, and $\Delta\pt$ and $\Delta y$ are the transverse
momentum and rapidity bin widths, respectively. The widths of the
$\pt$ bins are proportional to the $\pt$ resolution and thus increase
with \pt.

Because of the detector resolution and the steeply falling spectra,
the measured cross sections ($\stilde$) are smeared with respect to
the particle-level cross sections ($\sigma$).  Gaussian smearing
functions are obtained from the detector simulation and are used to
correct for the measured differences in the resolution between data
and simulation~\cite{Chatrchyan:2011ds}. These $\pt$-dependent resolutions are
folded with the NLO$\otimes$NP theory predictions, and are then used
to calculate the response matrices for jet \pt.  The unfolding is done
with the \RooUnfold package~\cite{ROOUNFOLD} using the D'Agostini
method~\cite{D'Agostini:1994zf}.  The unfolding reduces the measured cross
sections at $\abs{y} < 2.5$ ($2.5 \leq \abs{y} < 3.0$) by 5--20\% (15--30\%)
for $R=0.5$ and 5--25\% (15--40\%) for $R=0.7$. The large unfolding
factor at $2.5 \leq \abs{y} < 3.0$ is a consequence of the steep $\pt$
spectrum combined with the poor $\pt$ resolution in the region outside
the tracking coverage. The larger unfolding factor for $R=0.7$ than
for $R=0.5$ at $\pt<100\GeV$ is caused by the fact that jets with a
larger cone size are more affected by smearing from pileup.

The unfolding procedure is cross-checked against two alternative
methods. First, the NLO$\otimes$NP theory is smeared using the
smearing function and compared to the measured data. Second, the
\RooUnfold implementation of the singular-value decomposition (SVD)
method~\cite{Hocker:1995kb} is used to unsmear the data. All three
results (D'Agostini method, forward smearing, and SVD method) agree
within uncertainties.

\begin{figure*}[hbtp!]
  \centering
  \includegraphics[width=\textwidth]{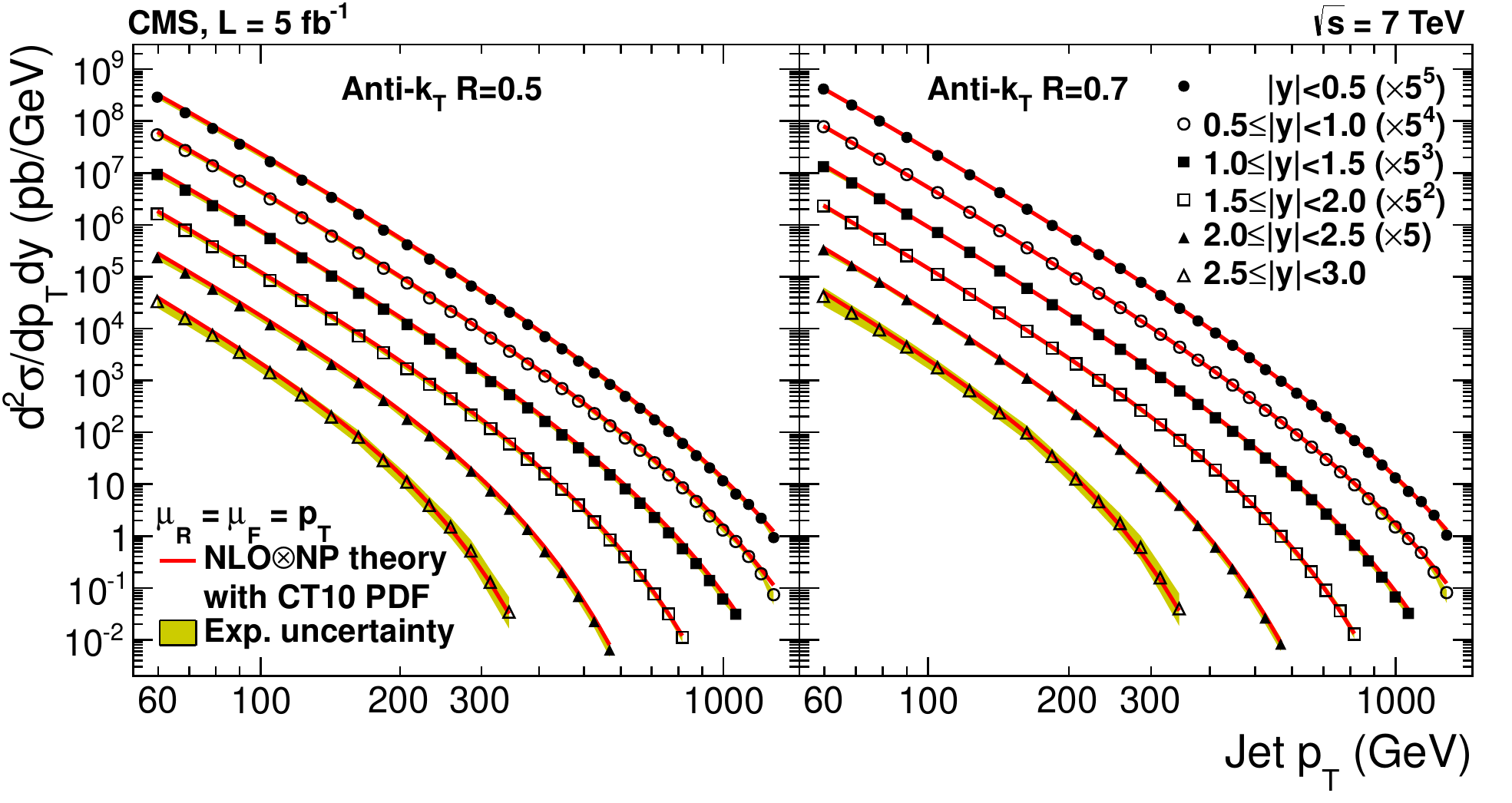}
  \caption[Cross section]
  {\label{fig:xsec}Unfolded inclusive jet cross section with anti-\kt
    $R=0.5$ (left) and 0.7 (right) compared to an NLO$\otimes$NP
    theory prediction using the CT10 NLO PDF set. The renormalization
    ($\mu_\mathrm{R}$) and factorization ($\mu_\mathrm{F}$) scales are
    defined to be the transverse momentum \pt of the jets.}
\end{figure*}

\begin{figure*}[hbtp!]
  \centering
  \includegraphics[width=\cmsFigWidth]{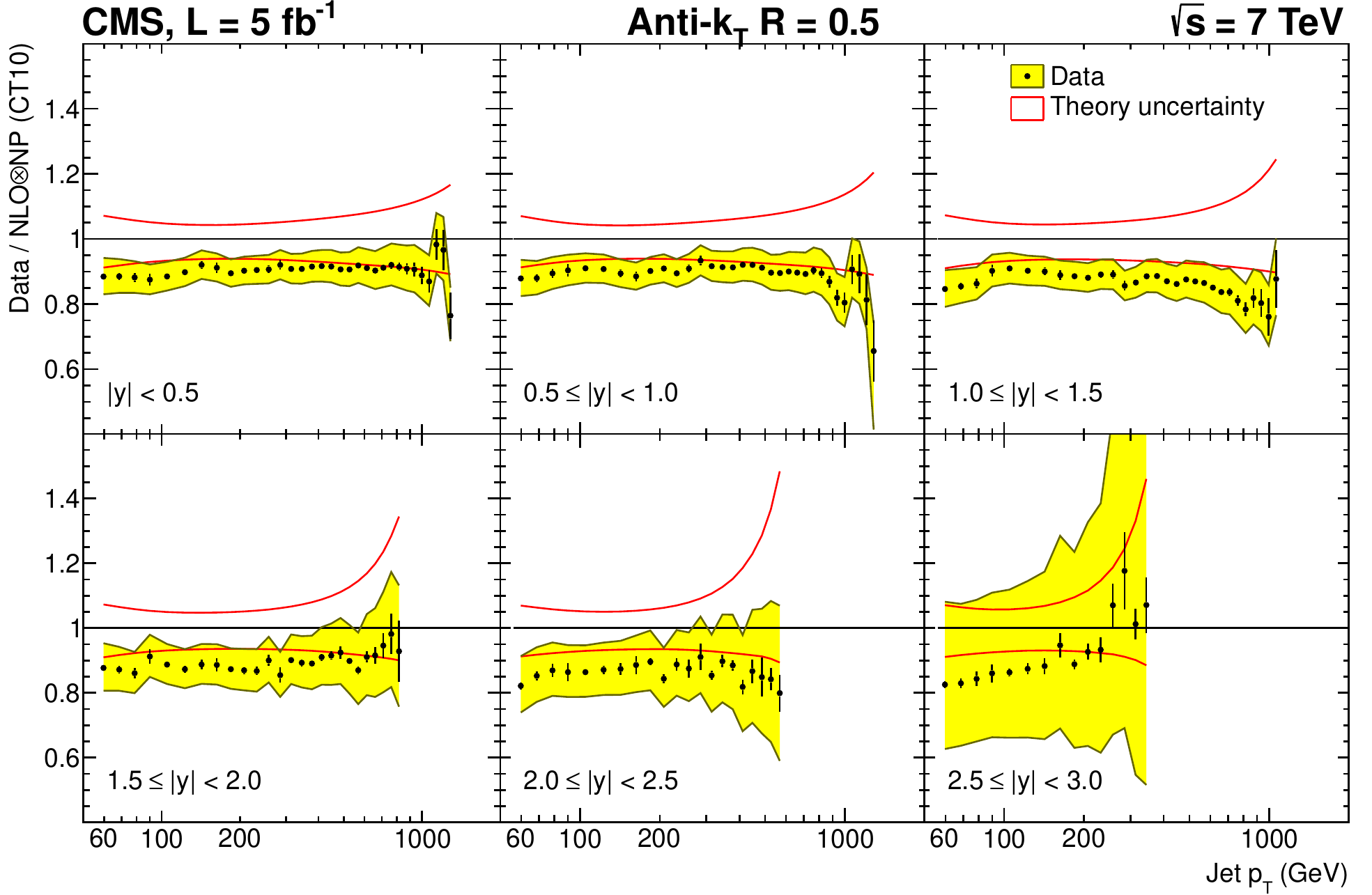}
  \includegraphics[width=\cmsFigWidth]{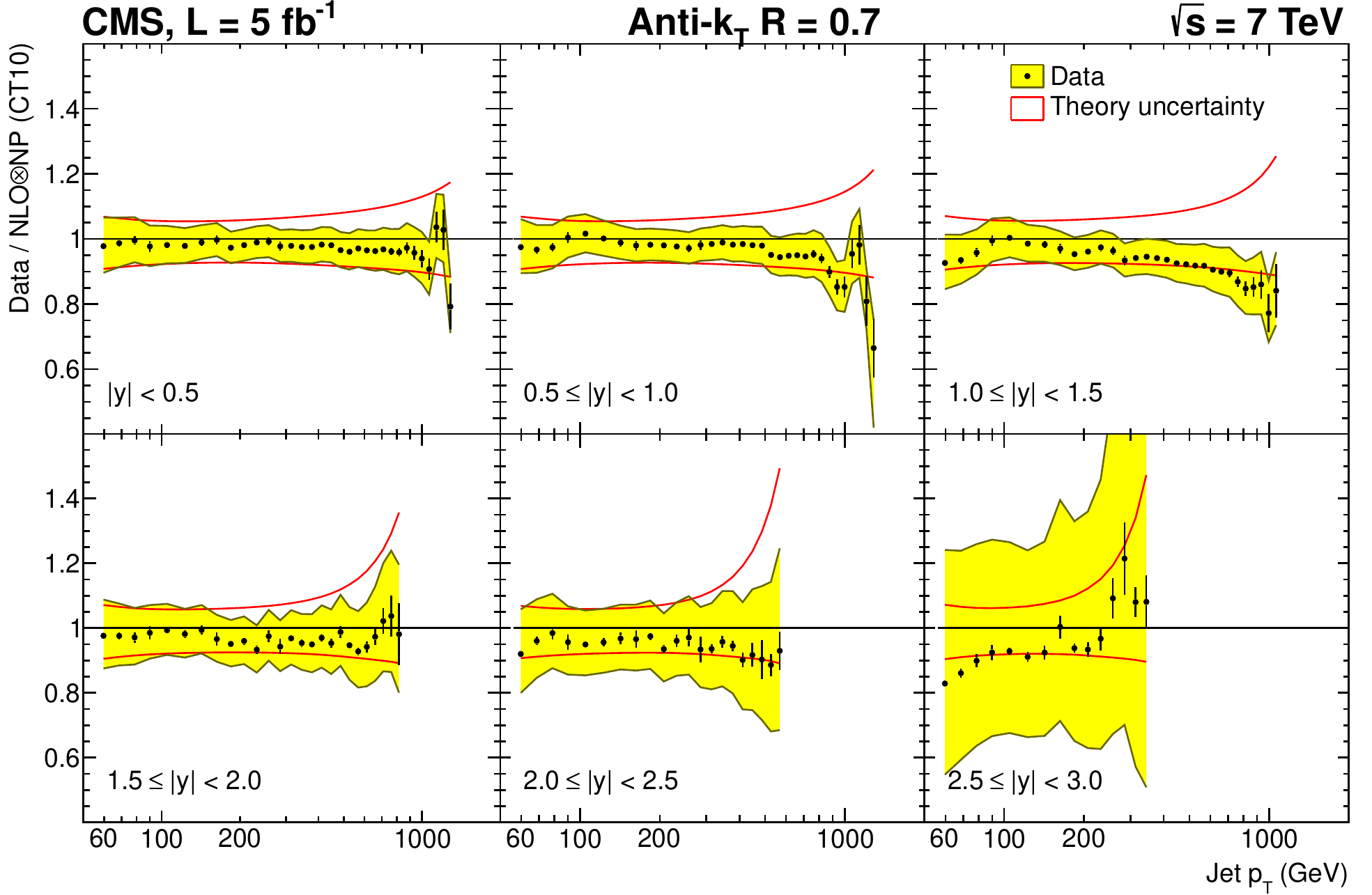}
  \caption[Cross section over theory]
  {\label{fig:xsecratio}Inclusive jet cross section with anti-\kt
    $R=0.5$ (top) and $R=0.7$ (bottom) divided by the NLO$\otimes$NP
    theory prediction using the CT10 NLO PDF set. The statistical and
    systematic uncertainties are represented by the error bars and the
    shaded band, respectively. The solid lines indicate the total
    theory uncertainty.
    The points with larger error bars occur at trigger boundaries.
  }
\end{figure*}

The unfolded inclusive jet cross section measurements with $R=0.5$ and
0.7 are shown in Fig.~\ref{fig:xsec}. Figure~\ref{fig:xsecratio}
shows the ratio of data to the NLO$\otimes$NP theory prediction using
the CT10 NLO PDF set~\cite{Lai:2010vv}.
The data agree with theory within uncertainties for both
jet radii. For $R=0.5$ the new measurements benefit from significantly
improved jet energy scale (JES) uncertainties compared to the previous
one~\cite{Chatrchyan:2011ab} and the much larger data sample used in this analysis increases
the number of jets available at high $\pt$. Contrarily, at low $\pt$
the larger single jet trigger prescales reduce the available number of
jets.  For $R=0.7$ the data set is identical to Ref.~\cite{Chatrchyan:2012bja}, but the measurement is
extended to lower $\pt$ and to higher rapidity. The total
uncertainties in this analysis are reduced with respect to the
previous one as discussed in Section~\ref{sec:pileup}.

The jet radius ratio, $\R(0.5,0.7)=\sigma_5/\sigma_7$, is obtained
from the bin-by-bin quotient of the unfolded cross sections,
$\sigma_5$ and $\sigma_7$, for $R=0.5$ and 0.7 respectively. The
statistical uncertainty is calculated separately to account for the
correlation between the two measurements. The details of the error
propagation are discussed in Appendix~\ref{appendixA}.

\subsection{Systematic uncertainties}

The main uncertainty sources and their 
impact is summarised in Tab.~\ref{table:Systematic}.
The dominant experimental uncertainties come from the subtraction of
the pileup offset in the JES correction and the jet $\pt$ resolution.
The total systematic uncertainty on $\R(0.5,0.7)$ varies from about
0.4\% at $\pt=1\TeV$ to 2\% at $\pt=60\GeV$ for $\abs{y} < 0.5$, and from
about 1.5\% at $\pt=600\GeV$ to 3.5\% at $\pt=60\GeV$ for $2.0 \leq
\abs{y} < 2.5$. Outside the tracker coverage at $2.5 \leq \abs{y} < 3.0$, the
uncertainty increases to between 3\% at $\pt=300\GeV$ and 8\% at
$\pt=60\GeV$.  The statistical uncertainties vary from a few per mil
to a couple of percent except at the highest $\pt$ (around the \TeV
scale), where they grow to 10\%.  The theory uncertainties amount
typically to 1 to 2\%, depending on the region. They are composed of
the scale dependence of the fixed-order perturbative calculations, of
the uncertainties in the PDFs, of the nonperturbative effects, and of
the statistical uncertainty in the cross section ratio prediction.

The luminosity uncertainty, which is relevant for the individual cross
section measurements, cancels out in the jet radius ratio, as do most
jet energy scale systematic uncertainties except for the pileup
corrections. The trigger efficiency, while almost negligible for
separate cross section measurements, becomes relevant for the jet
radius ratio when other larger systematic effects cancel out and the
correlations reduce the statistical uncertainty in the ratio. Other
sources of systematic uncertainty, such as the jet angular resolution,
are negligible.

The trigger efficiency uncertainty and the quadratic sum of all almost
negligible sources are assumed to be fully uncorrelated versus $\pt$
and $y$. The remaining sources are assumed to be fully correlated
versus $\pt$ and $y$ within three separate rapidity regions, but
uncorrelated between these regions: barrel ($\abs{y} < 1.5$), endcap ($1.5
\leq \abs{y} < 2.5$), and outside the tracking coverage ($2.5 \leq \abs{y} <
3.0$).

\begin{table}
\caption{\label{table:Systematic} Typical uncertainties on $\R(0.5,0.7)$.}    
\centering
\begin{scotch}{l | c  c} 
Uncertainty Source                   &  $\abs{y} < 2.5$ & $2.5 \leq \abs{y} < 3.0$ \\ \hline
Pileup       &    0.5--2\%     &          2--5\%       \\ 
Unfolding       &    1--2\%     &          5--7\%       \\ 
Trigger      & 0.5--1.5\%  & 0.5--1.5\%              \\ 
Statistical      &    0.2--10\%       &   0.2--10\%        \\ 
\end{scotch}
 
\end{table}

\subsubsection{Pileup uncertainty}
\label{sec:pileup}

The JES is the dominant source of systematic uncertainty for the
inclusive jet cross sections, but because the $R=0.5$ and 0.7 jets
are usually reconstructed with very similar $\pt$, the JES uncertainty
nearly cancels out in the ratio. A notable exception is the pileup
offset uncertainty, because the correction, and therefore the
uncertainty, is twice as large for the $R=0.7$ jets as for the $R=0.5$
jets.  The pileup uncertainty is the dominant systematic uncertainty
in this analysis over most of the phase space.

The JES pileup uncertainties cover differences in offset observed
between data and simulation, differences in the instantaneous
luminosity profile between the single jet triggers, and the $\stilde$
stability versus the instantaneous luminosity, which may indicate
residual pileup-dependent biases.  The earlier CMS
analysis~\cite{Chatrchyan:2011ab} also included JES uncertainties
based on simulation for the $\pt$ dependence of the offset and the
difference between the reconstructed offset and the true offset at
$\pt\sim30\GeV$. These uncertainties could be removed for the jet
radius ratio analysis because of improvements in the simulation.

The leading systematic uncertainty for $\abs{y} < 2.5$ is the stability of
$\stilde$ versus the instantaneous luminosity, while for $\abs{y} \geq 2.5$ the
differences between data and simulation are dominant.  The $\stilde$
stability uncertainty contributes 0.4--2\% at $\abs{y} < 0.5$ and 1--2\%
at $2.0 \leq \abs{y} < 3.0$, with the uncertainty increasing towards lower
$\pt$ and higher rapidity.  The data/MC differences contribute
0.5--1.5\% at $2.0 \leq \abs{y} < 2.5$ and 2--5\% at $2.5 \leq \abs{y} < 3.0$,
and increase towards low $\pt$. They are small or negligible for lower
rapidities.  Differences in the instantaneous luminosity profile
contribute less than about 0.5\% in the barrel at $\abs{y} < 1.5$, and are
about the same size as the data/MC differences in the endcaps within
tracker coverage at $1.5 \leq \abs{y} < 2.5$. Outside the tracker coverage
at $2.5 \leq \abs{y} < 3.0$ they contribute 1.0--2.5\%.

The uncertainty sources are assumed fully correlated between $R=0.5$
and 0.7, and are simultaneously propagated to the $R=0.5$ and
0.7 spectra before taking the jet radius ratio, one source at a
time.

\subsubsection{Unfolding uncertainty}

The unfolding correction depends on the jet energy resolution (JER)
and the $\pt$ spectrum slope. For the inclusive jet $\pt$ spectrum,
the relative JER uncertainty varies between 5\% and 15\% (30\%) for
$\abs{y} < 2.5$ ($2.5 \leq \abs{y} < 3.0$).

The JER uncertainty is propagated by smearing the NLO$\otimes$NP cross
section with smaller and larger values of the JER, and comparing the
resulting cross sections with the cross sections smeared with the
nominal JER\@. The relative JER uncertainty is treated as fully
correlated between $R=0.5$ and 0.7, and thus the uncertainty mostly
cancels for the jet radius ratio. Some residual uncertainty remains
mainly at $\pt<100\GeV$, where the magnitude of the JER differs
between $R=0.5$ and 0.7, because of additional smearing for the
larger cone size from the pileup offset.  The unfolding uncertainty at
$\pt=60\GeV$ varies between about 1\% for $\abs{y} < 0.5$, 2\% for $2.0
\leq \abs{y} < 2.5$, and 5--7\% for $2.5 \leq \abs{y} < 3.0$. It quickly
decreases to a sub-dominant uncertainty for $\pt=100\GeV$ and upwards,
and is practically negligible for $\pt>200\GeV$ in all rapidity bins.

\subsubsection{Trigger efficiency uncertainty}

The trigger turn-on curves for $R=0.7$ are less steep than for
$R=0.5$, which leads to relative inefficiencies near the trigger $\pt$
thresholds. The trigger efficiencies are estimated in simulation by
applying the trigger $\pt$ selections to $R=0.5$ jets measured in the
calorimeters, and comparing the results of a tag-and-probe
method~\cite{Khachatryan:2010xn} for data and MC.
The tag jet is required to have 100\% trigger efficiency, while the
unbiased PF probe jet is matched to a $R=0.5$ jet measured by the
calorimetric detectors to evaluate the trigger efficiency.
Differences between data and MC trigger efficiencies are at most
0.5--1.5\% and are taken as a systematic uncertainty, assumed to be
fully correlated between bins in $\pt$ and~$y$.

The maximum values of the trigger uncertainty are found near the steep
part of the trigger turn-on curves, which are also the bins with the
smallest statistical uncertainty. For the other bins the trigger
uncertainty is small or negligible compared to the statistical
uncertainty. Adding the trigger and the statistical contributions in
quadrature results in a total uncorrelated uncertainty of 0.5--2.0\%
for most $\pt$ bins, except at the highest $\pt$.

\subsubsection{Theory uncertainties in the NLO pQCD predictions}

The scale uncertainty due to the missing orders beyond NLO is
estimated with the conventional recipe of varying the renormalization
and factorization scales in the pQCD calculation for the cross section
ratio $\R(0.5,0.7)$. Six variations around the default choice of
$\mu_\mathrm{R} = \mu_\mathrm{F} = \pt$ for each jet are considered:
($\mu_\mathrm{R}/\pt$, $\mu_\mathrm{F}/\pt$) = (0.5, 0.5), (2, 2), (1, 0.5),
(1, 2), (0.5, 1), (2, 1). The maximal deviation of the six points is
considered as the total uncertainty.

The PDF uncertainty is evaluated by using the eigenvectors of the CT10
NLO PDF set~\cite{Lai:2010vv} for both cross sections, with $R = 0.5$
and 0.7. The total PDF uncertainty is propagated to
$\R(0.5,0.7)$ by considering it fully correlated between $R = 0.5$ and
0.7. The uncertainty induced by the strong coupling constant is
of the order of 1--2\% for individual cross sections and vanishes
nearly completely in the ratio.

The uncertainty caused by the modeling of nonperturbative effects is
estimated by taking half the difference of the \PYTHIAS and \HERWIGPP
predictions.

The scale uncertainty of the cross sections exceeds 5\% and can grow
up to 40\% in the forward region, but it cancels in the ratio and can
get as small as 1--2\%. It is, nevertheless, the overall dominant
theoretical uncertainty for the ratio analysis. Similarly, the PDF
uncertainty for the ratio is very small, while the NP uncertainty
remains important at low \pt, since it is sensitive to the difference
in jet area between $R=0.5$ and 0.7 jets.  Finally, the statistical
uncertainty of the theory prediction, which amounts to about 0.5\%,
does not cancel out in the ratio and it plays a role comparable to the
other sources.
\section{Results}%PLB 702(2011) 336-354

\begin{figure*}[hbtp!]
  \centering
  \includegraphics[width=\cmsFigWidth]{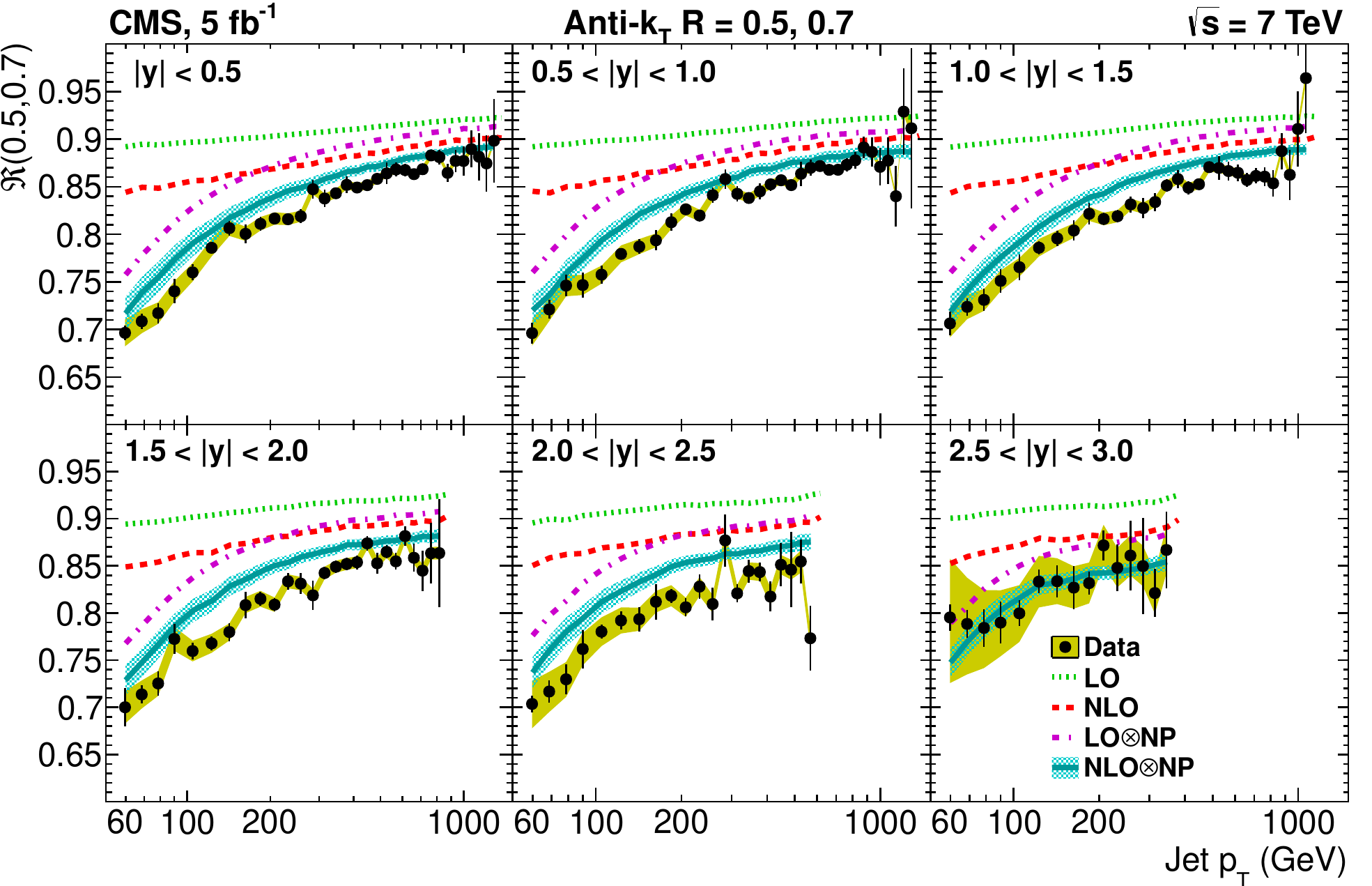}
  \includegraphics[width=\cmsFigWidth]{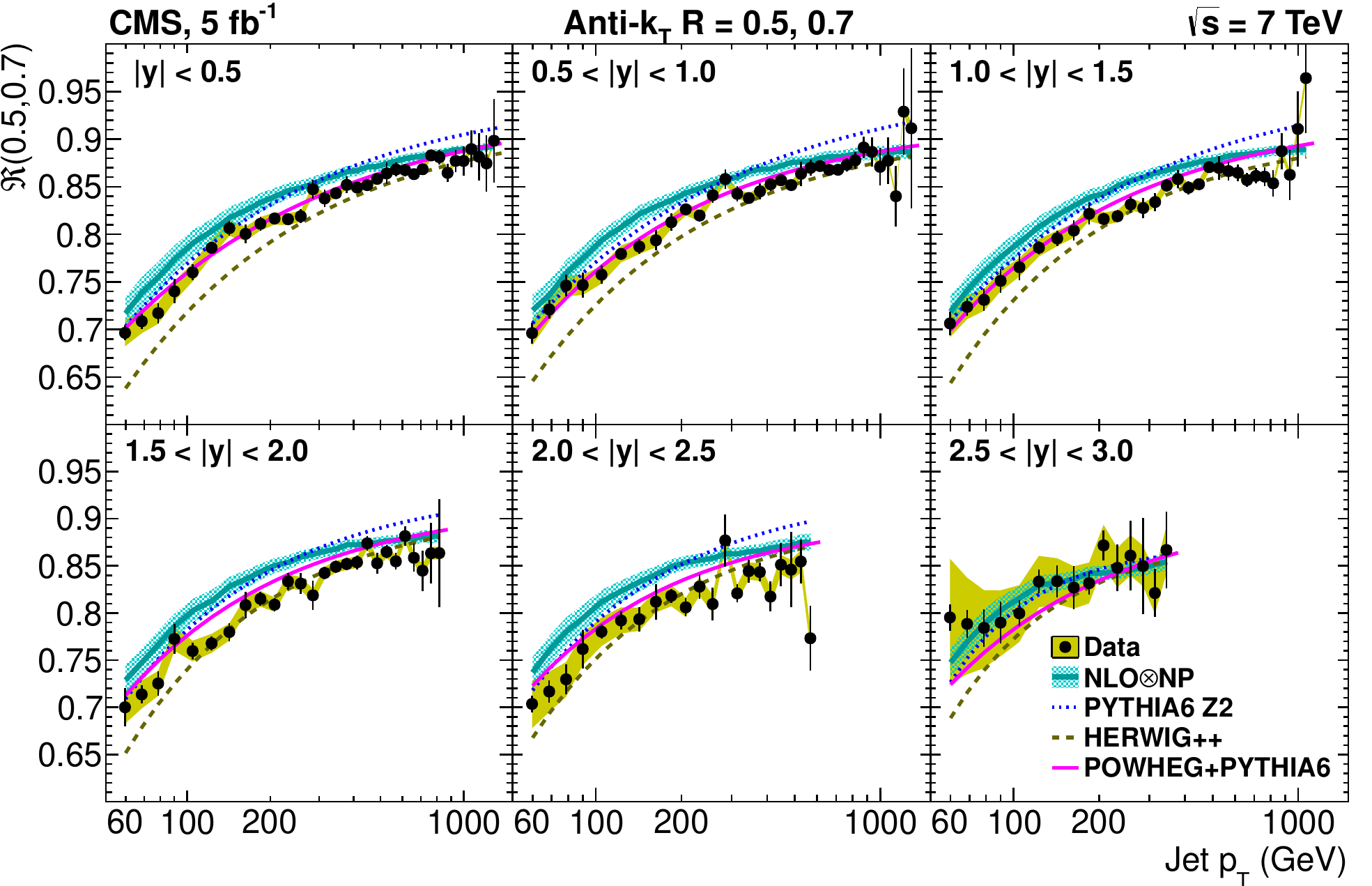}
  \caption[Results]
  {\label{fig:ak5ak7ratio}Jet radius ratio $\R(0.5,0.7)$ in six
    rapidity bins up to $\abs{y} = 3.0$, compared to LO and NLO with and
    without NP corrections (upper panel) and versus NLO$\otimes$NP and
    MC predictions (lower panel).
    The error bars on the data points represent the statistical and
    uncorrelated systematic uncertainty added in quadrature, and the
    shaded bands represent correlated systematic uncertainty.
    The NLO calculation was provided by G.~Soyez~\cite{Soyez:2011np}.
  }
\end{figure*}

The results for the jet radius ratio $\R(0.5,0.7)$ are presented for
all six bins of rapidity in Fig.~\ref{fig:ak5ak7ratio}. Each source of
systematic uncertainty is assumed to be fully correlated between the
$R=0.5$ and 0.7 cross section measurements, which is supported by
closure tests. Systematic uncertainties from the trigger efficiency
and a number of other small sources are considered as
uncorrelated
and are added in quadrature into a single uncorrelated systematic
source. The statistical uncertainty is propagated from the $R=0.5$ and
0.7 measurements taking into account the correlations induced by jet
reconstruction, dijet events, and unfolding. The uncorrelated
systematic uncertainty and the diagonal component of the statistical
uncertainty are added in quadrature for display purposes to give the
total uncorrelated uncertainty, as opposed to the correlated
systematic uncertainty.

In the central region, $\abs{y} < 2.5$, which benefits from the tracker
coverage, the systematic uncertainties are small and strongly
correlated between different $y$ bins. In contrast the forward region,
$2.5 \leq \abs{y} < 3.0$, relies mainly on the calorimeter information and
suffers from larger uncertainties. The central and forward regions are
uncorrelated in terms of systematic uncertainties.

The jet radius ratio does not exhibit a significant rapidity
dependence. The ratio rises toward unity with increasing $\pt$. From
the comparison to pQCD in the upper panel of
Fig.~\ref{fig:ak5ak7ratio} one concludes that in the inner rapidity
region of $\abs{y} < 2.5$, the theory is systematically above the data
with little rapidity dependence, while the NLO$\otimes$NP prediction
is closer to the data than the LO$\otimes$NP one. The pQCD predictions
without nonperturbative corrections are in clear disagreement with
the data. Nonperturbative effects are significant for $\pt < 1\TeV$,
but they are expected to be reliably estimated using the latest tunes
of \PYTHIAS and \HERWIGPP, for which the nonperturbative corrections
agree. Because of the much larger uncertainties in the outer rapidity
region with $2.5 \leq \abs{y} < 3.0$, no distinction between predictions
can be made except for pure LO and NLO, which also here lie
systematically above the data.

In the lower panel of Fig.~\ref{fig:ak5ak7ratio} the data are compared
to different Monte Carlo predictions. The best overall agreement is
provided by \POWHEG{}+\PYTHIAS. Comparing the parton showering
predictions of \PYTHIAS and \HERWIGPP to data exhibits agreement
across some regions of phase space, and disagreement in other
regions. The \PYTHIAS tune Z2 prediction agrees with data at the low
$\pt$ end of the measurement, where nonperturbative effects
dominate. This is where \PYTHIAS benefits most from having been tuned
to the LHC underlying event data. The \HERWIGPP predictions, on the
other hand, are in disagreement with the low $\pt$ data, which is
expected to be primarily due to the limitations of the underlying
event tune~2.3 in \HERWIGPP. This disagreement between the underlying
event in data and \HERWIGPP has been directly verified by observing
that for the same pileup conditions the energy density
$\rho$~\cite{Cacciari:2007fd} is larger by 0.3\GeV/rad$^2$ in
\HERWIGPP than in data, while \PYTHIAS describes well the energy
density in data. At higher $\pt$ the situation is reversed, with
\HERWIGPP describing the data and \PYTHIAS disagreeing. This fact
might be related to the better ability of \HERWIGPP to describe the
high-$\pt$ jet substructure with respect to
\PYTHIAS~\cite{Chatrchyan:2012ypy}.

\section{Summary}

The inclusive jet cross section has been measured for two different
jet radii, $R = 0.5$ and 0.7, as a function of the jet rapidity $y$
and transverse momentum $\pt$. Special care has been taken to fully
account for correlations when the jet radius ratio $\R(0.5,0.7)$ is
derived from these measurements. 
Although the cross sections
themselves can be described within the theoretical and experimental uncertainties 
by predictions of
pQCD at NLO (including terms up to $\alps^3$), this is not
the case for the ratio $\R(0.5,0.7)$. The cancellation of systematic
uncertainties in the ratio poses a more stringent test of the theoretical
predictions than the individual cross section measurements do.
For this three-jet observable
$\R(0.5,0.7)$, which looks in detail into the pattern of QCD
radiation, NLO (including terms up to $\alps^4$), even when
complemented with nonperturbative corrections, is in clear
disagreement with the data. This is not unexpected, since at most four
partons are available at this order to characterize any $R$
dependence.

The MC event generators \PYTHIAS and \HERWIGPP, which rely on parton
showers to describe three-jet observables, are in better accord with
the measured jet radius ratio $\R(0.5,0.7)$ than the fixed-order
predictions. The best description of this ratio is obtained by
matching the cross section prediction at NLO with parton showers, as
studied here
using
\POWHEG with \PYTHIAS for the
showering, underlying event, and hadronization parts. The observations
above hold for all regions with $\abs{y} < 2.5$, while for $\abs{y} \geq 2.5$
the experimental uncertainty limits the ability to discriminate
between different predictions.

In summary, it has been demonstrated that jet radius $R$ dependent
effects, measurable in data, require pQCD predictions with at least
one order higher than NLO or a combination of NLO cross sections
matched to parton shower models to be sufficiently characterized by
theory.

\begin{acknowledgments}
\hyphenation{Bundes-ministerium Forschungs-gemeinschaft Forschungs-zentren}  We would like to thank G.~Soyez for providing us
with the NLO predictions for the jet radius ratio. We congratulate our colleagues in the CERN accelerator departments for the excellent performance of the LHC and thank the technical and administrative staffs at CERN and at other CMS institutes for their contributions to the success of the CMS effort. In addition, we gratefully acknowledge the computing centres and personnel of the Worldwide LHC Computing Grid for delivering so effectively the computing infrastructure essential to our analyses. Finally, we acknowledge the enduring support for the construction and operation of the LHC and the CMS detector provided by the following funding agencies: the Austrian Federal Ministry of Science, Research and Economy and the Austrian Science Fund; the Belgian Fonds de la Recherche Scientifique, and Fonds voor Wetenschappelijk Onderzoek; the Brazilian Funding Agencies (CNPq, CAPES, FAPERJ, and FAPESP); the Bulgarian Ministry of Education and Science; CERN; the Chinese Academy of Sciences, Ministry of Science and Technology, and National Natural Science Foundation of China; the Colombian Funding Agency (COLCIENCIAS); the Croatian Ministry of Science, Education and Sport, and the Croatian Science Foundation; the Research Promotion Foundation, Cyprus; the Ministry of Education and Research, Estonian Research Council via IUT23-4 and IUT23-6 and European Regional Development Fund, Estonia; the Academy of Finland, Finnish Ministry of Education and Culture, and Helsinki Institute of Physics; the Institut National de Physique Nucl\'eaire et de Physique des Particules~/~CNRS, and Commissariat \`a l'\'Energie Atomique et aux \'Energies Alternatives~/~CEA, France; the Bundesministerium f\"ur Bildung und Forschung, Deutsche Forschungsgemeinschaft, and Helmholtz-Gemeinschaft Deutscher Forschungszentren, Germany; the General Secretariat for Research and Technology, Greece; the National Scientific Research Foundation, and National Innovation Office, Hungary; the Department of Atomic Energy and the Department of Science and Technology, India; the Institute for Studies in Theoretical Physics and Mathematics, Iran; the Science Foundation, Ireland; the Istituto Nazionale di Fisica Nucleare, Italy; the Korean Ministry of Education, Science and Technology and the World Class University program of NRF, Republic of Korea; the Lithuanian Academy of Sciences; the Ministry of Education, and University of Malaya (Malaysia); the Mexican Funding Agencies (CINVESTAV, CONACYT, SEP, and UASLP-FAI); the Ministry of Business, Innovation and Employment, New Zealand; the Pakistan Atomic Energy Commission; the Ministry of Science and Higher Education and the National Science Centre, Poland; the Funda\c{c}\~ao para a Ci\^encia e a Tecnologia, Portugal; JINR, Dubna; the Ministry of Education and Science of the Russian Federation, the Federal Agency of Atomic Energy of the Russian Federation, Russian Academy of Sciences, and the Russian Foundation for Basic Research; the Ministry of Education, Science and Technological Development of Serbia; the Secretar\'{\i}a de Estado de Investigaci\'on, Desarrollo e Innovaci\'on and Programa Consolider-Ingenio 2010, Spain; the Swiss Funding Agencies (ETH Board, ETH Zurich, PSI, SNF, UniZH, Canton Zurich, and SER); the Ministry of Science and Technology, Taipei; the Thailand Center of Excellence in Physics, the Institute for the Promotion of Teaching Science and Technology of Thailand, Special Task Force for Activating Research and the National Science and Technology Development Agency of Thailand; the Scientific and Technical Research Council of Turkey, and Turkish Atomic Energy Authority; the National Academy of Sciences of Ukraine, and State Fund for Fundamental Researches, Ukraine; the Science and Technology Facilities Council, UK; the US Department of Energy, and the US National Science Foundation.

Individuals have received support from the Marie-Curie programme and the European Research Council and EPLANET (European Union); the Leventis Foundation; the A. P. Sloan Foundation; the Alexander von Humboldt Foundation; the Belgian Federal Science Policy Office; the Fonds pour la Formation \`a la Recherche dans l'Industrie et dans l'Agriculture (FRIA-Belgium); the Agentschap voor Innovatie door Wetenschap en Technologie (IWT-Belgium); the Ministry of Education, Youth and Sports (MEYS) of the Czech Republic; the Council of Science and Industrial Research, India; the HOMING PLUS programme of Foundation for Polish Science, cofinanced from European Union, Regional Development Fund; the Compagnia di San Paolo (Torino); and the Thalis and Aristeia programmes cofinanced by EU-ESF and the Greek NSRF.
\end{acknowledgments}

\bibliography{auto_generated}   % will be created by the tdr script.

\appendix
\section{Error propagation}
\label{appendixA}

The procedure of extracting from data the jet radius ratio
$\R(0.5,0.7)$
and its covariance matrix
consists of the following steps: the data are in the form of exclusive
jet radius-pair production cross sections $m_{x,pq}^{ij}$,
$m_{5,pq}^{ij}$, $m_{7,pq}^{ij}$, for jet radius pairs
$(R=0.5,0.7)$, $(R=0.5,0.5)$, and $(R=0.7,0.7)$, respectively,
with given number $q$ and $p$ of jets in $\pt$ bins with indices $i$
and $j$, respectively.  From these the inclusive jet cross sections
$\stilde_5$ and $\stilde_7$ are extracted as functions of $\pt$, using
\begin{equation}\begin{aligned}
  \stilde_{5,i} &= \sum_{p,q} p\cdot m_{5,pq}^{ij} = \sum_{p,q} p\cdot m_{x,pq}^{ij}, \quad\text{for any $j$},\\
  \stilde_{7,j} &= \sum_{p,q} q\cdot m_{7,pq}^{ij} = \sum_{p,q} q\cdot m_{x,pq}^{ij}, \quad \text{for any $i$}.
\end{aligned}\end{equation}
As a result of unfolding, $\stilde_5$ and $\stilde_7$ are converted into
particle-level cross sections $\sigma_5$ and $\sigma_7$, from which
the jet radius ratio $\R(0.5,0.7)$ is computed for each $\pt$ bin.

The error propagation can be summarized in matrix notation:
\begin{equation}\begin{aligned}
  \label{eq:mW}
  W_{55,ij} &=  \sum_{p,q} pq\cdot \var{m_{5,pq}^{ij}},\\
  W_{77,ij} &= \sum_{p,q} pq\cdot \var{m_{7,pq}^{ij}}, \\
  W_{57,ij} &= \sum_{p,q} pq\cdot \var{m_{x,pq}^{ij}}, \\
\end{aligned}
\end{equation}
\begin{equation}
  \label{eq:mB}
\left.\begin{aligned}
  B_{5,ij} &=  \frac{\partial{\sigma_{5,i}}}{\partial{\stilde_{5,j}}},\\
  B_{7,ij} &= \frac{\partial{\sigma_{7,i}}}{\partial{\stilde_{7,j}}},
  \end{aligned}\right.\qquad\text{(evaluated numerically)}
\end{equation}
\begin{equation}\begin{aligned}
  V_{55} &=  B_5 W_{55} B_5^T,\\
  V_{77} &= B_7 W_{77} B_7^T,\\
  V_{57} &= B_5 W_{57} B_7^T,
  \label{eq:mV}
\end{aligned}\end{equation}

giving
\begin{align}
\label{eq:mVa}
V &=  \left[\begin{array}{ll}
      \hphantom{(}V_{55}   & V_{57}\\
      (V_{57})^T & V_{77} \\
    \end{array}\right],
\\
\label{eq:mA}
A_{ik} &=
  \left\{\begin{array}{ll}
      \R_i\frac{1}{\sigma_{5,i}} & \text{if $k = i$, and $i\leq n$},\\
      -\R_i\frac{1}{\sigma_{7,i}} & \text{if $k = i+n$, and $i\leq n$},\\
      0 & \text{otherwise},
    \end{array}\right.\\
\label{eq:mU}
  U &=  A V A^T.
\end{align}

The $W$ matrices in Eq.~(\ref{eq:mW}) give the correlations of the jet
cross sections in the various $\pt$ bins, for $(R=0.5,0.5)$,
$(R=0.7,0.7)$, and $(R=0.5,0.7)$ jets; the correlations in the
first two arise from dijet events, and the correlations in the last
one primarily from the fact that a single jet can appear in both
$R=0.5$ and 0.7 categories.  Most of the jets are reconstructed
with both $R=0.5$ and 0.7 clustering parameters, and often fall in
the same $(\pt,y)$ bin.  The measured correlation between
$\stilde_{5,i}$ and $\stilde_{7,j}$ for bin $i=j$ in data is about 0.4 at
$\pt=50$\GeV, rising to 0.65 at $\pt=100$\GeV, and finally to 0.85 at
$\pt \ge 1$\TeV. The correlation is almost independent of rapidity for
a fixed $\pt$.  At low $\pt$ there is fairly strong correlation of up
to 0.4 between bins $i=j-1$ and $j$, and of up to 0.1 between bins
$i=j-2$ and $j$.  A small correlation of up to 0.1 between bins
$i=j+1$ and $j$ is also observed at high $\pt$ at $\abs{y}<1.0$ because of
dijet events contributing to adjacent $\pt$ bins. This correlation is
also present for jets reconstructed with the same radius parameter,
and is considered in the error propagation.  The correlation between
other bins is negligible and only bin pairs coming from the same
single-jet trigger are considered correlated.

The $B$ matrices in Eq.~(\ref{eq:mB}) transform the covariance
matrices $W$ of the measured spectra $\stilde_5$ and $\stilde_7$ to the
covariance matrices $V$ for the unfolded spectra $\sigma_5$ and
$\sigma_7$.  Equations~(\ref{eq:mV}) and (\ref{eq:mU}) follow from
standard error propagation, as
in Eq.~(1.55) of Ref.~\cite{cowan1998statistical}.  The partial derivatives
$\partial\sigma_i/\partial \stilde_j$ in Eq.~(\ref{eq:mB}) are evaluated
by numerically differentiating the D'Agostini unfolding, where the
$\sigma_{5,i}$ and $\sigma_{7,i}$ are the unfolded cross sections,
$\stilde_{5,i}$ and $\stilde_{7,j}$ are the corresponding smeared cross
sections, and $\R_i=\sigma_{5,i}/\sigma_{7,i}$ is the jet radius
ratio.  The matrices $V_{55}$ and $V_{77}$ agree to within 10\% of
those returned by \RooUnfold for $R=0.5$ and 0.7 $\pt$ spectra,
respectively, but also account for the bin-to-bin correlations induced
by dijet events.

For the purposes of error propagation, the $\stilde_5$ and $\stilde_7$
data are represented as a single $2n$ vector with $\stilde_5$ at indices
1 to $n$ and $\stilde_7$ at indices $n+1$ to $2n$.  The matrix $V$ in
Eq.~(\ref{eq:mVa}) therefore has dimensions of $2n \times 2n$ and the
matrix $A$ in Eq.~(\ref{eq:mA}) has dimensions $n\times 2n$.

Finally, the covariance matrix $U$ in Eq.~(\ref{eq:mU}) for the jet
radius ratio $\R(0.5,0.7)$ is calculated using the error propagation
matrix $A$ and the combined covariance matrix $V$ for the unfolded jet
cross sections with $R=0.5$ and 0.7.

\begin{figure*}[hbt]
  \centering
  \includegraphics[width=0.49\textwidth]{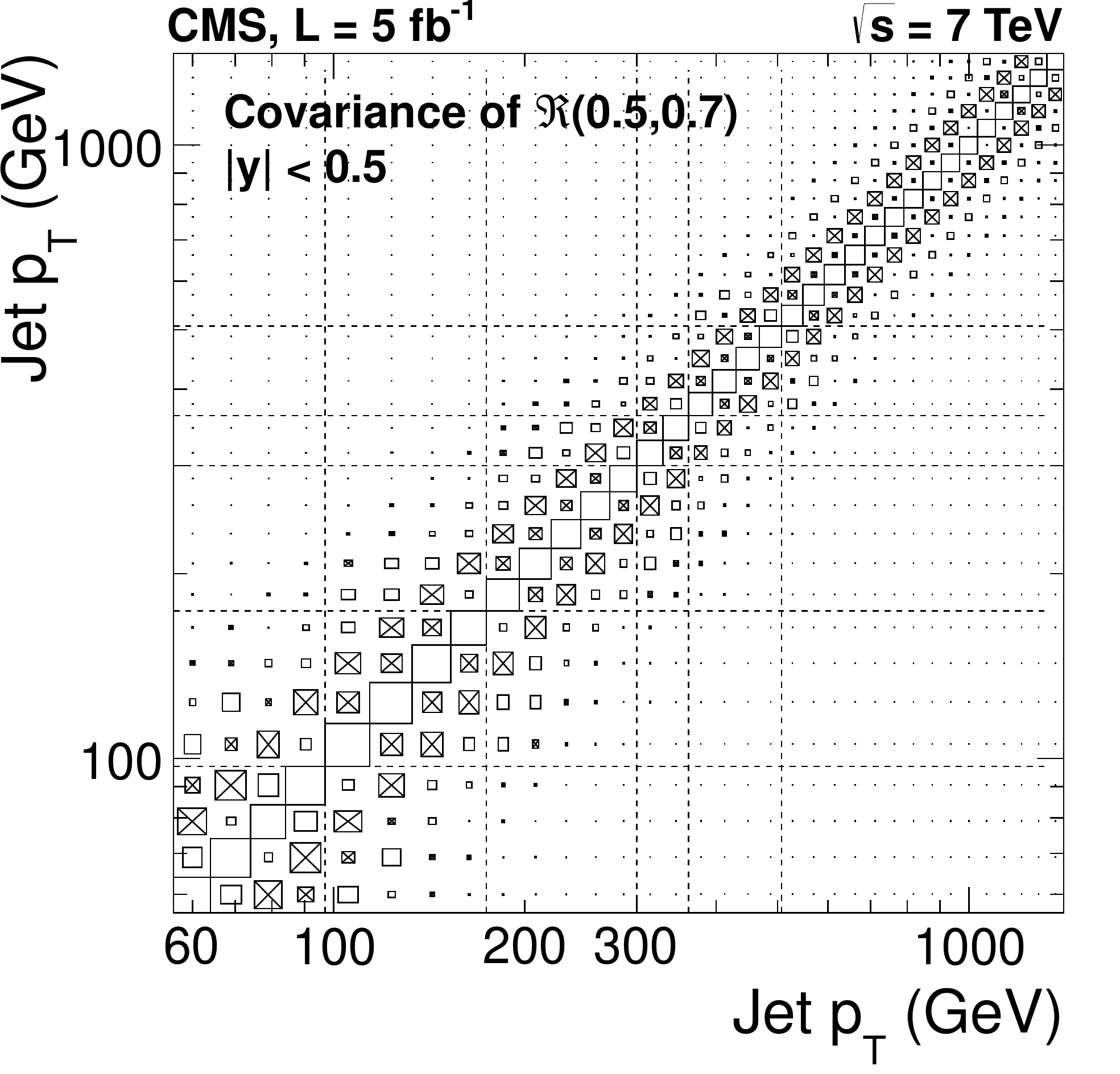}
  \includegraphics[width=0.49\textwidth]{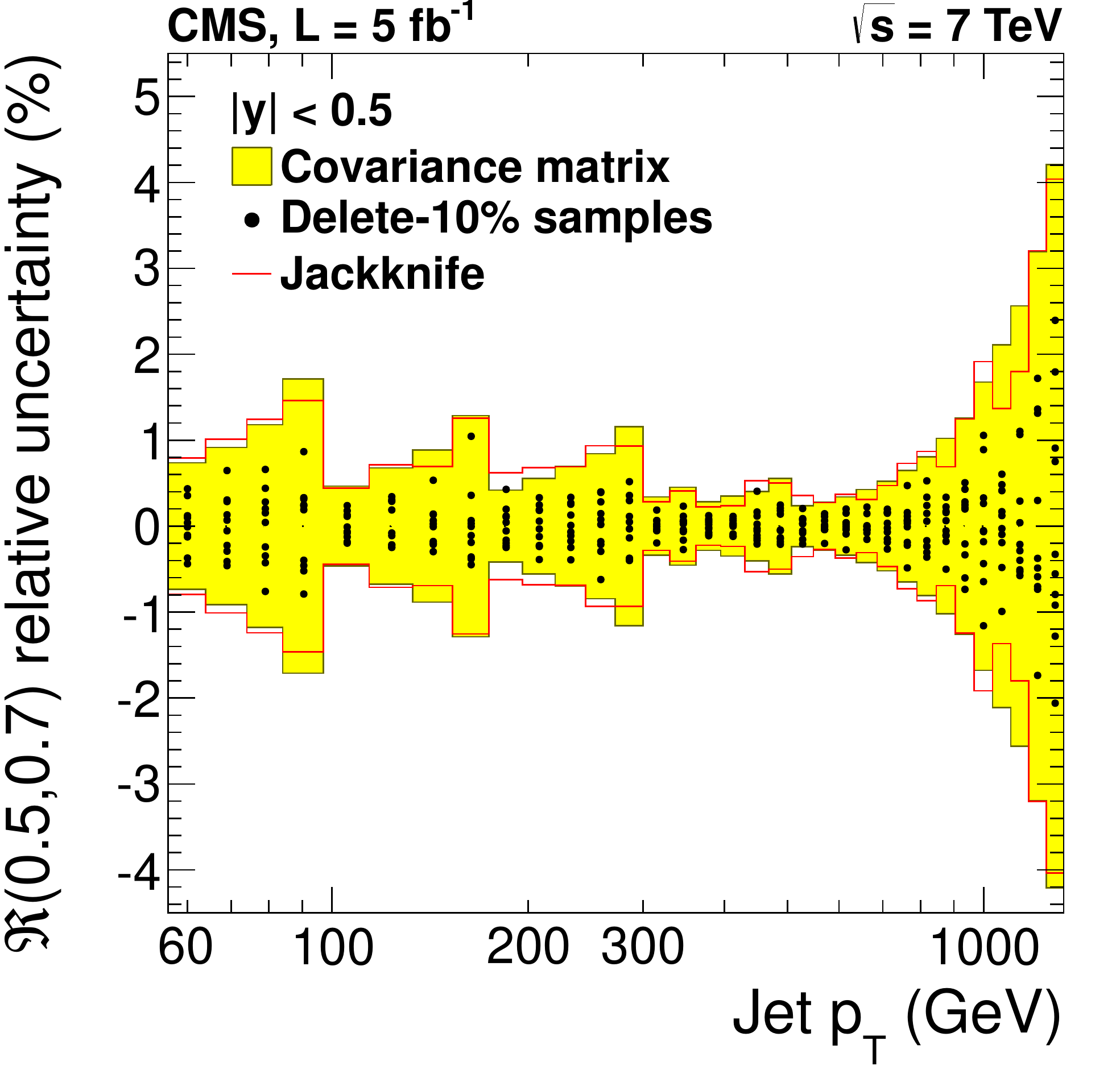}
  \caption[Covariance matrix]
  {\label{fig:cov} (Left) Covariance matrix $U$ for the jet radius
    ratio $\R(0.5,0.7)$, normalized by the diagonal elements to show
    the level of correlation.  Dashed horizontal and vertical lines
    indicate the analysis $\pt$ thresholds corresponding to different
    triggers.  The size of the boxes relative to bin size is
    proportional to the correlation coefficient in the range from -1
    to 1. The diagonal elements are 1 and thus indicative of the
    variable bin size.  The crossed boxes corresponds to
    anticorrelation, while the open boxes correspond to positive
    correlations between two bins.  (Right) Comparison of the square
    root of the covariance matrix diagonals with a random sampling
    estimate using the delete-$d$ ($d=10$\%) jackknife method. The
    differences between the full data set and the ten delete-$d$
    samples are shown by the full circles.}
\end{figure*}

The resulting covariance matrix $U$ is shown in Fig.~\ref{fig:cov}
(left) for $\abs{y} < 0.5$.  The strong anticorrelation observed between
neighboring bins is similar to that observed for individual spectra,
and is mainly an artifact of the D'Agostini unfolding.  The
statistical uncertainty for each bin of $\R(0.5,0.7)$ is illustrated
as the square root of the corresponding diagonal element of the
covariance matrix in Fig.~\ref{fig:cov} (right).  Given the relative
complexity of the error propagation, the statistical uncertainties are
validated using a variant of bootstrap methods called the delete-$d$
jackknife \cite{efron1993introduction}.  In this method the data are
divided into ten samples, each having a non-overlapping uniformly
distributed fraction $d=10$\% of the events removed.  The ten sets of
jet cross-sections are used to obtain a covariance matrix, which is
scaled by $(1-d)/d=9$ to estimate the (co)variance of the original
sample.  The variances obtained by error propagation agree with the
jackknife estimate in all rapidity bins within the expected jackknife
uncertainty.

\cleardoublepage \section{The CMS Collaboration \label{app:collab}}\begin{sloppypar}\hyphenpenalty=5000\widowpenalty=500\clubpenalty=5000\input{SMP-13-002-authorlist.tex}\end{sloppypar}
\end{document}

%% file: SMP-13-002-authorlist.tex
\textbf{Yerevan Physics Institute,  Yerevan,  Armenia}\\*[0pt]
S.~Chatrchyan, V.~Khachatryan, A.M.~Sirunyan, A.~Tumasyan
\vskip\cmsinstskip
\textbf{Institut f\"{u}r Hochenergiephysik der OeAW,  Wien,  Austria}\\*[0pt]
W.~Adam, T.~Bergauer, M.~Dragicevic, J.~Er\"{o}, C.~Fabjan\cmsAuthorMark{1}, M.~Friedl, R.~Fr\"{u}hwirth\cmsAuthorMark{1}, V.M.~Ghete, C.~Hartl, N.~H\"{o}rmann, J.~Hrubec, M.~Jeitler\cmsAuthorMark{1}, W.~Kiesenhofer, V.~Kn\"{u}nz, M.~Krammer\cmsAuthorMark{1}, I.~Kr\"{a}tschmer, D.~Liko, I.~Mikulec, D.~Rabady\cmsAuthorMark{2}, B.~Rahbaran, H.~Rohringer, R.~Sch\"{o}fbeck, J.~Strauss, A.~Taurok, W.~Treberer-Treberspurg, W.~Waltenberger, C.-E.~Wulz\cmsAuthorMark{1}
% ...
\vskip\cmsinstskip
% ...
1:~~Also at Vienna University of Technology, Vienna, Austria\\
2:~~Also at CERN, European Organization for Nuclear Research, Geneva, Switzerland\\
% ...


%% file: SMP-13-002_temp.bbl
\providecommand{\href}[2]{#2}\begingroup\raggedright\begin{thebibliography}{10}%
\makeatletter
\providecommand{\hrefCMSnoop }[0]{\@secondoftwo}%
\makeatother
\providecommand{\doi}{\texttt{doi:}\begingroup \urlstyle{tt}\Url}

\bibitem{Ellis:1990ek}
\hrefCMSnoop {} {S.~D. Ellis, Z.~Kunszt, and D.~E. Soper, ``One-Jet Inclusive
  Cross Section at Order $\alpha_s^3$: Quarks and Gluons'',} \textit{ Phys.
  Rev. Lett.} \textbf{ 64} (1990) 2121,
\href{http://dx.doi.org/10.1103/PhysRevLett.64.2121}{\doi{10.1103/PhysRevLett.64.2121}}.
%%CITATION = PRLTA,64,2121;%%.

\bibitem{Ellis:1992en}
\hrefCMSnoop {} {S.~D. Ellis, Z.~Kunszt, and D.~E. Soper, ``Two-Jet Production
  in Hadron Collisions at Order $\alpha_s^3$ in {QCD}'',} \textit{ Phys. Rev.
  Lett.} \textbf{ 69} (1992) 1496,
\href{http://dx.doi.org/10.1103/PhysRevLett.69.1496}{\doi{10.1103/PhysRevLett.69.1496}}.
%%CITATION = PRLTA,69,1496;%%.

\bibitem{Giele:1993dj}
\hrefCMSnoop {} {W.~T. Giele, E.~W.~N. Glover, and D.~A. Kosower, ``Higher
  order corrections to jet cross-sections in hadron colliders'',} \textit{
  Nucl. Phys. B} \textbf{ 403} (1993) 633,
  \href{http://dx.doi.org/10.1016/0550-3213(93)90365-V}{\doi{10.1016/0550-3213(93)90365-V}},
\href{http://www.arXiv.org/abs/hep-ph/9302225}{\texttt{ arXiv:hep-ph/9302225}}.
%%CITATION = HEP-PH/9302225;%%.

\bibitem{Currie:2013dwa}
\hrefCMSnoop {} {J.~Currie, A.~Gehrmann-De~Ridder, E.~W.~N. Glover, and
  J.~Pires, ``{NNLO} {QCD} corrections to jet production at hadron colliders
  from gluon scattering'',} \textit{ J. High Ener. Phys.} \textbf{ 01} (2014)
  110,
  \href{http://dx.doi.org/10.1007/JHEP01(2014)110}{\doi{10.1007/JHEP01(2014)110}},
\href{http://www.arXiv.org/abs/1310.3993}{\texttt{ arXiv:1310.3993}}.
%%CITATION = ARXIV:1310.3993;%%.

\bibitem{Cacciari:2008gp}
\hrefCMSnoop {} {M.~Cacciari, G.~P. Salam, and G.~Soyez, ``The {anti-$k_t$} jet
  clustering algorithm'',} \textit{ J. High Ener. Phys.} \textbf{ 04} (2008)
  063,
  \href{http://dx.doi.org/10.1088/1126-6708/2008/04/063}{\doi{10.1088/1126-6708/2008/04/063}},
\href{http://www.arXiv.org/abs/0802.1189}{\texttt{ arXiv:0802.1189}}.
%%CITATION = 0802.1189;%%.

\bibitem{Catani:1991hj}
S.~Catani\hrefCMSnoop {} { {et~al.}, ``New clustering algorithm for multijet
  cross-sections in e$^{+}$ e$^{-}$ annihilation'',} \textit{ Phys. Lett. B}
  \textbf{ 269} (1991) 432,
\href{http://dx.doi.org/10.1016/0370-2693(91)90196-W}{\doi{10.1016/0370-2693(91)90196-W}}.
%%CITATION = PHLTA,B269,432;%%.

\bibitem{Brown:1991hx}
\hrefCMSnoop {} {N.~Brown and W.~J. Stirling, ``Finding jets and summing soft
  gluons: A New algorithm'',} \textit{ Z. Phys. C} \textbf{ 53} (1992) 629,
\href{http://dx.doi.org/10.1007/BF01559740}{\doi{10.1007/BF01559740}}.
%%CITATION = ZEPYA,C53,629;%%.

\bibitem{Catani:1992zp}
\hrefCMSnoop {} {S.~Catani, Y.~L. Dokshitzer, and B.~R. Webber, ``The
  $k_\perp$-clustering algorithm for jets in deep inelastic scattering and
  hadron collisions'',} \textit{ Phys. Lett. B} \textbf{ 285} (1992) 291,
\href{http://dx.doi.org/10.1016/0370-2693(92)91467-N}{\doi{10.1016/0370-2693(92)91467-N}}.
%%CITATION = PHLTA,B285,291;%%.

\bibitem{Ellis:1993tq}
\hrefCMSnoop {} {S.~D. Ellis and D.~E. Soper, ``Successive combination jet
  algorithm for hadron collisions'',} \textit{ Phys. Rev. D} \textbf{ 48}
  (1993) 3160,
  \href{http://dx.doi.org/10.1103/PhysRevD.48.3160}{\doi{10.1103/PhysRevD.48.3160}},
\href{http://www.arXiv.org/abs/hep-ph/9305266}{\texttt{ arXiv:hep-ph/9305266}}.
%%CITATION = HEP-PH/9305266;%%.

\bibitem{Affolder:2001fa}
\hrefCMSnoop {} {{ CDF} Collaboration, ``Measurement of the inclusive jet cross
  section in $\bar{p}p$ collisions at $\sqrt{s} = 1.8$ {TeV}'',} \textit{ Phys.
  Rev. D} \textbf{ 64} (2001) 032001,
  \href{http://dx.doi.org/10.1103/PhysRevD.64.032001}{\doi{10.1103/PhysRevD.64.032001}},
  \href{http://www.arXiv.org/abs/hep-ph/0102074}{\texttt{
  arXiv:hep-ph/0102074}}.
See also Erratum,
  \href{http://dx.doi.org/10.1103/PhysRevD.65.039903}{\texttt{doi:10.1103/PhysRevD.65.039903}},
  Phys. Rev. D 65, 039903 (2002).
%%CITATION = HEP-PH/0102074;%%.

\bibitem{Abulencia:2007ez}
\hrefCMSnoop {} {{ CDF} Collaboration, ``Measurement of the Inclusive Jet Cross
  Section using the $k_{\rm T}$ algorithm in $p\overline{p}$ Collisions at
  $\sqrt{s}$ = 1.96 {TeV} with the {CDF II} Detector'',} \textit{ Phys. Rev. D}
  \textbf{ 75} (2007) 092006,
  \href{http://dx.doi.org/10.1103/PhysRevD.75.092006}{\doi{10.1103/PhysRevD.75.092006}},
  \href{http://www.arXiv.org/abs/hep-ex/0701051}{\texttt{
  arXiv:hep-ex/0701051}}.
See also Publisher's Note,
  \href{http://dx.doi.org/10.1103/PhysRevD.75.119901}{\texttt{doi:10.1103/PhysRevD.75.119901}},
  Phys. Rev. D 75, 119901 (2007).
%%CITATION = HEP-EX/0701051;%%.

\bibitem{Aaltonen:2008eq}
\hrefCMSnoop {} {{ CDF} Collaboration, ``Measurement of the Inclusive Jet Cross
  Section at the Fermilab Tevatron ${p\bar{p}}$ Collider Using a Cone-Based Jet
  Algorithm'',} \textit{ Phys. Rev. D} \textbf{ 78} (2008) 052006,
  \href{http://dx.doi.org/10.1103/PhysRevD.78.052006}{\doi{10.1103/PhysRevD.78.052006}},
  \href{http://www.arXiv.org/abs/0807.2204}{\texttt{ arXiv:0807.2204}}.
See also Erratum,
  \href{http://dx.doi.org/10.1103/PhysRevD.79.119902}{\texttt{doi:10.1103/PhysRevD.79.119902}},
  Phys. Rev. D 79, 119902 (2009).
%%CITATION = ARXIV:0807.2204;%%.

\bibitem{Abbott:2000ew}
\hrefCMSnoop {} {{ D0} Collaboration, ``Inclusive jet production in $p\bar{p}$
  collisions'',} \textit{ Phys. Rev. Lett.} \textbf{ 86} (2001) 1707,
  \href{http://dx.doi.org/10.1103/PhysRevLett.86.1707}{\doi{10.1103/PhysRevLett.86.1707}},
\href{http://www.arXiv.org/abs/hep-ex/0011036}{\texttt{ arXiv:hep-ex/0011036}}.
%%CITATION = HEP-EX/0011036;%%.

\bibitem{Abazov:2001hb}
\hrefCMSnoop {} {{ D0} Collaboration, ``The inclusive jet cross-section in
  $p\bar{p}$ collisions at $\sqrt{s} = 1.8$ TeV using the $k_T$ algorithm'',}
  \textit{ Phys. Lett. B} \textbf{ 525} (2002) 211,
  \href{http://dx.doi.org/10.1016/S0370-2693(01)01441-1}{\doi{10.1016/S0370-2693(01)01441-1}},
\href{http://www.arXiv.org/abs/hep-ex/0109041}{\texttt{ arXiv:hep-ex/0109041}}.
%%CITATION = HEP-EX/0109041;%%.

\bibitem{Abazov:2011vi}
\hrefCMSnoop {} {{ D0} Collaboration, ``Measurement of the inclusive jet cross
  section in $p \bar {p}$ collisions at $\sqrt{s}=1.96$ TeV'',} \textit{ Phys.
  Rev. D} \textbf{ 85} (2012) 052006,
  \href{http://dx.doi.org/10.1103/PhysRevD.85.052006}{\doi{10.1103/PhysRevD.85.052006}},
\href{http://www.arXiv.org/abs/1110.3771}{\texttt{ arXiv:1110.3771}}.
%%CITATION = ARXIV:1110.3771;%%.

\bibitem{Aad:2011fc}
\hrefCMSnoop {} {{ ATLAS} Collaboration, ``Measurement of inclusive jet and
  dijet production in $pp$ collisions at $\sqrt{s}=7$ {TeV} using the {ATLAS}
  detector'',} \textit{ Phys. Rev. D} \textbf{ 86} (2012) 014022,
  \href{http://dx.doi.org/10.1103/PhysRevD.86.014022}{\doi{10.1103/PhysRevD.86.014022}},
\href{http://www.arXiv.org/abs/1112.6297}{\texttt{ arXiv:1112.6297}}.
%%CITATION = ARXIV:1112.6297;%%.

\bibitem{Aad:2013lpa}
\hrefCMSnoop {} {{ ATLAS} Collaboration, ``Measurement of the inclusive jet
  cross section in $pp$ collisions at $\sqrt{s}$=2.76 {TeV} and comparison to
  the inclusive jet cross section at $\sqrt{s}$=7 {TeV} using the {ATLAS}
  detector'',} \textit{ Eur. Phys. J. C} \textbf{ 73} (2013) 2509,
  \href{http://dx.doi.org/10.1140/epjc/s10052-013-2509-4}{\doi{10.1140/epjc/s10052-013-2509-4}},
\href{http://www.arXiv.org/abs/1304.4739}{\texttt{ arXiv:1304.4739}}.
%%CITATION = ARXIV:1304.4739;%%.

\bibitem{Chatrchyan:2011ab}
\hrefCMSnoop {} {{ CMS} Collaboration, ``Measurement of the Inclusive Jet Cross
  Section in {pp} Collisions at $\sqrt{s}=7$ {TeV}'',} \textit{ Phys. Rev.
  Lett.} \textbf{ 107} (2011) 132001,
  \href{http://dx.doi.org/10.1103/PhysRevLett.107.132001}{\doi{10.1103/PhysRevLett.107.132001}},
\href{http://www.arXiv.org/abs/1106.0208}{\texttt{ arXiv:1106.0208}}.
%%CITATION = ARXIV:1106.0208;%%.

\bibitem{Chatrchyan:2012gw}
\hrefCMSnoop {} {{ CMS} Collaboration, ``Measurement of the inclusive
  production cross sections for forward jets and for dijet events with one
  forward and one central jet in {pp} collisions at $\sqrt{s}=7$ {TeV}'',}
  \textit{ J. High Ener. Phys.} \textbf{ 06} (2012) 036,
  \href{http://dx.doi.org/10.1007/JHEP06(2012)036}{\doi{10.1007/JHEP06(2012)036}},
\href{http://www.arXiv.org/abs/1202.0704}{\texttt{ arXiv:1202.0704}}.
%%CITATION = ARXIV:1202.0704;%%.

\bibitem{Chatrchyan:2012bja}
\hrefCMSnoop {} {{ CMS} Collaboration, ``Measurements of differential jet cross
  sections in proton-proton collisions at $\sqrt{s}=7$ {TeV} with the {CMS}
  detector'',} \textit{ Phys. Rev. D} \textbf{ 87} (2013) 112002,
  \href{http://dx.doi.org/10.1103/PhysRevD.87.112002}{\doi{10.1103/PhysRevD.87.112002}},
\href{http://www.arXiv.org/abs/1212.6660}{\texttt{ arXiv:1212.6660}}.
%%CITATION = ARXIV:1212.6660;%%.

\bibitem{Abelev:2013fn}
\hrefCMSnoop {} {{ ALICE} Collaboration, ``Measurement of the inclusive
  differential jet cross section in $pp$ collisions at $\sqrt{s} = 2.76$
  {TeV}'',} \textit{ Phys. Lett. B} \textbf{ 722} (2013) 262,
  \href{http://dx.doi.org/10.1016/j.physletb.2013.04.026}{\doi{10.1016/j.physletb.2013.04.026}},
\href{http://www.arXiv.org/abs/1301.3475}{\texttt{ arXiv:1301.3475}}.
%%CITATION = ARXIV:1301.3475;%%.

\bibitem{Dasgupta:2007wa}
\hrefCMSnoop {} {M.~Dasgupta, L.~Magnea, and G.~P. Salam, ``Non-perturbative
  {QCD} effects in jets at hadron colliders'',} \textit{ J. High Ener. Phys.}
  \textbf{ 02} (2008) 055,
  \href{http://dx.doi.org/10.1088/1126-6708/2008/02/055}{\doi{10.1088/1126-6708/2008/02/055}},
\href{http://www.arXiv.org/abs/0712.3014}{\texttt{ arXiv:0712.3014}}.
%%CITATION = 0712.3014;%%.

\bibitem{Cacciari:2008gd}
\hrefCMSnoop {} {M.~Cacciari, J.~Rojo, G.~P. Salam, and G.~Soyez, ``Quantifying
  the performance of jet definitions for kinematic reconstruction at the
  {LHC}'',} \textit{ J. High Ener. Phys.} \textbf{ 12} (2008) 032,
  \href{http://dx.doi.org/10.1088/1126-6708/2008/12/032}{\doi{10.1088/1126-6708/2008/12/032}},
\href{http://www.arXiv.org/abs/0810.1304}{\texttt{ arXiv:0810.1304}}.
%%CITATION = 0810.1304;%%.

\bibitem{Nagy:2001fj}
\hrefCMSnoop {} {Z.~Nagy, ``Three-jet cross sections in hadron-hadron
  collisions at next-to-leading order'',} \textit{ Phys. Rev. Lett.} \textbf{
  88} (2002) 122003,
  \href{http://dx.doi.org/10.1103/PhysRevLett.88.122003}{\doi{10.1103/PhysRevLett.88.122003}},
\href{http://www.arXiv.org/abs/hep-ph/0110315}{\texttt{ arXiv:hep-ph/0110315}}.
%%CITATION = HEP-PH/0110315;%%.

\bibitem{Nagy:2003tz}
\hrefCMSnoop {} {Z.~Nagy, ``Next-to-leading order calculation of three-jet
  observables in hadron-hadron collisions'',} \textit{ Phys. Rev. D} \textbf{
  68} (2003) 094002,
  \href{http://dx.doi.org/10.1103/PhysRevD.68.094002}{\doi{10.1103/PhysRevD.68.094002}},
\href{http://www.arXiv.org/abs/hep-ph/0307268}{\texttt{ arXiv:hep-ph/0307268}}.
%%CITATION = HEP-PH/0307268;%%.

\bibitem{Soyez:2011np}
\hrefCMSnoop {} {G.~Soyez, ``A Simple description of jet cross-section
  ratios'',} \textit{ Phys. Lett. B} \textbf{ 698} (2011) 59,
  \href{http://dx.doi.org/10.1016/j.physletb.2011.02.061}{\doi{10.1016/j.physletb.2011.02.061}},
\href{http://www.arXiv.org/abs/1101.2665}{\texttt{ arXiv:1101.2665}}.
%%CITATION = ARXIV:1101.2665;%%.

\bibitem{Sjostrand:2006za}
\hrefCMSnoop {} {T.~Sj{\"o}strand, S.~Mrenna, and P.~Z. Skands, ``PYTHIA 6.4
  Physics and Manual'',} \textit{ J. High Ener. Phys.} \textbf{ 05} (2006) 026,
  \href{http://dx.doi.org/10.1088/1126-6708/2006/05/026}{\doi{10.1088/1126-6708/2006/05/026}},
\href{http://www.arXiv.org/abs/hep-ph/0603175}{\texttt{ arXiv:hep-ph/0603175}}.
%%CITATION = HEP-PH/0603175;%%.

\bibitem{Bahr:2008pv}
M.~B{\"a}hr\hrefCMSnoop {} { {et~al.}, ``Herwig++ physics and manual'',}
  \textit{ Eur. Phys. J. C} \textbf{ 58} (2008) 639,
  \href{http://dx.doi.org/10.1140/epjc/s10052-008-0798-9}{\doi{10.1140/epjc/s10052-008-0798-9}},
\href{http://www.arXiv.org/abs/0803.0883}{\texttt{ arXiv:0803.0883}}.
%%CITATION = 0803.0883;%%.

\bibitem{Alioli:2010xa}
S.~Alioli\hrefCMSnoop {} { {et~al.}, ``Jet pair production in {POWHEG}'',}
  \textit{ J. High Ener. Phys.} \textbf{ 04} (2011) 081,
  \href{http://dx.doi.org/10.1007/JHEP04(2011)081}{\doi{10.1007/JHEP04(2011)081}},
\href{http://www.arXiv.org/abs/1012.3380}{\texttt{ arXiv:1012.3380}}.
%%CITATION = ARXIV:1012.3380;%%.

\bibitem{Abramowicz:2010ke}
\hrefCMSnoop {} {{ ZEUS} Collaboration, ``Inclusive-jet cross sections in {NC}
  {DIS} at {HERA} and a comparison of the {kT}, {anti-kT} and {SIScone} jet
  algorithms'',} \textit{ Phys. Lett. B} \textbf{ 691} (2010) 127,
  \href{http://dx.doi.org/10.1016/j.physletb.2010.06.015}{\doi{10.1016/j.physletb.2010.06.015}},
\href{http://www.arXiv.org/abs/1003.2923}{\texttt{ arXiv:1003.2923}}.
%%CITATION = ARXIV:1003.2923;%%.

\bibitem{Chatrchyan:2008aa}
\hrefCMSnoop {} {{ CMS} Collaboration, ``The {CMS} experiment at the {CERN}
  {LHC}'',} \textit{ J. Inst.} \textbf{ 3} (2008) S08004,
\href{http://dx.doi.org/10.1088/1748-0221/3/08/S08004}{\doi{10.1088/1748-0221/3/08/S08004}}.
%%CITATION = JINST,3,S08004;%%.

\bibitem{CMS-PAS-PFT-09-001}
\href {http://cdsweb.cern.ch/record/1194487} {{ CMS} Collaboration,
  ``Particle--Flow Event Reconstruction in {CMS} and Performance for Jets,
  Taus, and {\MET}'',} CMS Physics Analysis Summary CMS-PAS-PFT-09-001, 2009.

\bibitem{Cacciari:2011ma}
\hrefCMSnoop {} {M.~Cacciari, G.~P. Salam, and G.~Soyez, ``{FastJet} User
  Manual'',} \textit{ Eur. Phys. J. C} \textbf{ 72} (2012) 1896,
  \href{http://dx.doi.org/10.1140/epjc/s10052-012-1896-2}{\doi{10.1140/epjc/s10052-012-1896-2}},
\href{http://www.arXiv.org/abs/1111.6097}{\texttt{ arXiv:1111.6097}}.
%%CITATION = ARXIV:1111.6097;%%.

\bibitem{Chatrchyan:2011ds}
\hrefCMSnoop {} {{ CMS} Collaboration, ``Determination of Jet Energy
  Calibration and Transverse Momentum Resolution in {CMS}'',} \textit{ J.
  Inst.} \textbf{ 6} (2011) P11002,
  \href{http://dx.doi.org/10.1088/1748-0221/6/11/P11002}{\doi{10.1088/1748-0221/6/11/P11002}},
\href{http://www.arXiv.org/abs/1107.4277}{\texttt{ arXiv:1107.4277}}.
%%CITATION = ARXIV:1107.4277;%%.

\bibitem{Cacciari:2007fd}
\hrefCMSnoop {} {M.~Cacciari and G.~P. Salam, ``Pileup subtraction using jet
  areas'',} \textit{ Phys. Lett. B} \textbf{ 659} (2008) 119,
  \href{http://dx.doi.org/10.1016/j.physletb.2007.09.077}{\doi{10.1016/j.physletb.2007.09.077}},
\href{http://www.arXiv.org/abs/0707.1378}{\texttt{ arXiv:0707.1378}}.
%%CITATION = 0707.1378;%%.

\bibitem{Bengtsson:1986et}
\hrefCMSnoop {} {M.~Bengtsson and T.~Sj{\"o}strand, ``A Comparative Study of
  Coherent and Noncoherent Parton Shower Evolution'',} \textit{ Nucl. Phys. B}
  \textbf{ 289} (1987) 810,
\href{http://dx.doi.org/10.1016/0550-3213(87)90407-X}{\doi{10.1016/0550-3213(87)90407-X}}.
%%CITATION = NUPHA,B289,810;%%.

\bibitem{Bengtsson:1986hr}
\hrefCMSnoop {} {M.~Bengtsson and T.~Sj{\"o}strand, ``Coherent Parton Showers
  Versus Matrix Elements: Implications of {PETRA} - {PEP} Data'',} \textit{
  Phys. Lett. B} \textbf{ 185} (1987) 435,
\href{http://dx.doi.org/10.1016/0370-2693(87)91031-8}{\doi{10.1016/0370-2693(87)91031-8}}.
%%CITATION = PHLTA,B185,435;%%.

\bibitem{Sjostrand:2004ef}
\hrefCMSnoop {} {T.~Sj{\"o}strand and P.~Z. Skands,
  ``Transverse-momentum-ordered showers and interleaved multiple
  interactions'',} \textit{ Eur. Phys. J. C} \textbf{ 39} (2005) 129,
  \href{http://dx.doi.org/10.1140/epjc/s2004-02084-y}{\doi{10.1140/epjc/s2004-02084-y}},
\href{http://www.arXiv.org/abs/hep-ph/0408302}{\texttt{ arXiv:hep-ph/0408302}}.
%%CITATION = HEP-PH/0408302;%%.

\bibitem{Sjostrand:1987su}
\hrefCMSnoop {} {T.~Sj{\"o}strand and M.~van Zijl, ``A Multiple Interaction
  Model for the Event Structure in Hadron Collisions'',} \textit{ Phys. Rev. D}
  \textbf{ 36} (1987) 2019,
\href{http://dx.doi.org/10.1103/PhysRevD.36.2019}{\doi{10.1103/PhysRevD.36.2019}}.
%%CITATION = PHRVA,D36,2019;%%.

\bibitem{Sjostrand:2004pf}
\hrefCMSnoop {} {T.~Sj{\"o}strand and P.~Z. Skands, ``Multiple interactions and
  the structure of beam remnants'',} \textit{ J. High Ener. Phys.} \textbf{ 03}
  (2004) 053,
  \href{http://dx.doi.org/10.1088/1126-6708/2004/03/053}{\doi{10.1088/1126-6708/2004/03/053}},
\href{http://www.arXiv.org/abs/hep-ph/0402078}{\texttt{ arXiv:hep-ph/0402078}}.
%%CITATION = HEP-PH/0402078;%%.

\bibitem{Andersson:1983ia}
\hrefCMSnoop {} {B.~Andersson, G.~Gustafson, G.~Ingelman, and T.~Sj{\"o}strand,
  ``Parton Fragmentation and String Dynamics'',} \textit{ Phys. Rept.} \textbf{
  97} (1983) 31,
\href{http://dx.doi.org/10.1016/0370-1573(83)90080-7}{\doi{10.1016/0370-1573(83)90080-7}}.
%%CITATION = PRPLC,97,31;%%.

\bibitem{Andersson:1983jt}
\hrefCMSnoop {} {B.~Andersson, G.~Gustafson, and B.~S{\"o}derberg, ``A General
  Model for Jet Fragmentation'',} \textit{ Z. Phys. C} \textbf{ 20} (1983) 317,
\href{http://dx.doi.org/10.1007/BF01407824}{\doi{10.1007/BF01407824}}.
%%CITATION = ZEPYA,C20,317;%%.

\bibitem{Sjostrand:1984iu}
\hrefCMSnoop {} {T.~Sj{\"o}strand, ``The merging of jets'',} \textit{ Phys.
  Lett. B} \textbf{ 142} (1984) 420,
\href{http://dx.doi.org/10.1016/0370-2693(84)91354-6}{\doi{10.1016/0370-2693(84)91354-6}}.
%%CITATION = PHLTA,B142,420;%%.

\bibitem{tunesZ2}
\hrefCMSnoop {} {R.~Field, ``Early LHC Underlying Event Data - Findings and
  Surprises'',} (2010).
\href{http://www.arXiv.org/abs/1010.3558}{\texttt{ arXiv:1010.3558}}.
%%CITATION = ARXIV:1010.3558;%%.

\bibitem{Pumplin:2002vw}
J.~Pumplin\hrefCMSnoop {} { {et~al.}, ``{New generation of parton distributions
  with uncertainties from global QCD analysis}'',} \textit{ J. High Ener.
  Phys.} \textbf{ 07} (2002) 012,
  \href{http://dx.doi.org/10.1088/1126-6708/2002/07/012}{\doi{10.1088/1126-6708/2002/07/012}},
  \href{http://www.arXiv.org/abs/hep-ph/0201195}{\texttt{
  arXiv:hep-ph/0201195}}.

\bibitem{Marchesini:1987cf}
\hrefCMSnoop {} {G.~Marchesini and B.~R. Webber, ``Monte Carlo simulation of
  general hard processes with coherent {QCD} radiation'',} \textit{ Nucl. Phys.
  B} \textbf{ 310} (1988) 461,
\href{http://dx.doi.org/10.1016/0550-3213(88)90089-2}{\doi{10.1016/0550-3213(88)90089-2}}.
%%CITATION = NUPHA,B310,461;%%.

\bibitem{Gieseke:2003rz}
\hrefCMSnoop {} {S.~Gieseke, P.~Stephens, and B.~Webber, ``New formalism for
  {QCD} parton showers'',} \textit{ J. High Ener. Phys.} \textbf{ 12} (2003)
  045,
  \href{http://dx.doi.org/10.1088/1126-6708/2003/12/045}{\doi{10.1088/1126-6708/2003/12/045}},
\href{http://www.arXiv.org/abs/hep-ph/0310083}{\texttt{ arXiv:hep-ph/0310083}}.
%%CITATION = HEP-PH/0310083;%%.

\bibitem{Bahr:2008dy}
\hrefCMSnoop {} {M.~B{\"a}hr, S.~Gieseke, and M.~H. Seymour, ``Simulation of
  multiple partonic interactions in {Herwig++}'',} \textit{ J. High Ener.
  Phys.} \textbf{ 07} (2008) 076,
  \href{http://dx.doi.org/10.1088/1126-6708/2008/07/076}{\doi{10.1088/1126-6708/2008/07/076}},
\href{http://www.arXiv.org/abs/0803.3633}{\texttt{ arXiv:0803.3633}}.
%%CITATION = 0803.3633;%%.

\bibitem{Webber:1983if}
\hrefCMSnoop {} {B.~R. Webber, ``A {QCD} model for jet fragmentation including
  soft gluon interference'',} \textit{ Nucl. Phys. B} \textbf{ 238} (1984) 492,
\href{http://dx.doi.org/10.1016/0550-3213(84)90333-X}{\doi{10.1016/0550-3213(84)90333-X}}.
%%CITATION = NUPHA,B238,492;%%.

\bibitem{Nason:2004rx}
\hrefCMSnoop {} {P.~Nason, ``A New method for combining {NLO} {QCD} with shower
  Monte Carlo algorithms'',} \textit{ J. High Ener. Phys.} \textbf{ 11} (2004)
  040,
  \href{http://dx.doi.org/10.1088/1126-6708/2004/11/040}{\doi{10.1088/1126-6708/2004/11/040}},
\href{http://www.arXiv.org/abs/hep-ph/0409146}{\texttt{ arXiv:hep-ph/0409146}}.
%%CITATION = HEP-PH/0409146;%%.

\bibitem{Frixione:2007vw}
\hrefCMSnoop {} {S.~Frixione, P.~Nason, and C.~Oleari, ``Matching {NLO} {QCD}
  computations with Parton Shower simulations: the {POWHEG} method'',} \textit{
  J. High Ener. Phys.} \textbf{ 11} (2007) 070,
  \href{http://dx.doi.org/10.1088/1126-6708/2007/11/070}{\doi{10.1088/1126-6708/2007/11/070}},
\href{http://www.arXiv.org/abs/0709.2092}{\texttt{ arXiv:0709.2092}}.
%%CITATION = ARXIV:0709.2092;%%.

\bibitem{Alioli:2010xd}
\hrefCMSnoop {} {S.~Alioli, P.~Nason, C.~Oleari, and E.~Re, ``A general
  framework for implementing {NLO} calculations in shower Monte Carlo programs:
  the {POWHEG} {BOX}'',} \textit{ J. High Ener. Phys.} \textbf{ 06} (2010) 043,
  \href{http://dx.doi.org/10.1007/JHEP06(2010)043}{\doi{10.1007/JHEP06(2010)043}},
\href{http://www.arXiv.org/abs/1002.2581}{\texttt{ arXiv:1002.2581}}.
%%CITATION = 1002.2581;%%.

\bibitem{Britzger:2012bs}
\hrefCMSnoop {} {D.~Britzger, K.~Rabbertz, F.~Stober, and M.~Wobisch, ``New
  features in version 2 of the {fastNLO} project'',} in \textit{ Proceedings of
  the XX Int. Workshop on Deep-Inelastic Scattering and Related Subjects
  (DIS2012)}, p.~217.
\newblock 2012.
\newblock \href{http://www.arXiv.org/abs/1208.3641}{\texttt{ arXiv:1208.3641}}.
\newblock
\href{http://dx.doi.org/10.3204/DESY-PROC-2012-02/165}{\doi{10.3204/DESY-PROC-2012-02/165}}.
%%CITATION = ARXIV:1208.3641;%%.

\bibitem{ROOUNFOLD}
\hrefCMSnoop {} {T.~Adye, ``Unfolding algorithms and tests using
  {RooUnfold}'',} (2011).
\href{http://www.arXiv.org/abs/1105.1160}{\texttt{ arXiv:1105.1160}}.
%%CITATION = ARXIV:1105.1160;%%.

\bibitem{D'Agostini:1994zf}
\hrefCMSnoop {} {G.~D'Agostini, ``A Multidimensional unfolding method based on
  Bayes' theorem'',} \textit{ Nucl. Instrum. Meth. A} \textbf{ 362} (1995) 487,
\href{http://dx.doi.org/10.1016/0168-9002(95)00274-X}{\doi{10.1016/0168-9002(95)00274-X}}.
%%CITATION = NUIMA,A362,487;%%.

\bibitem{Hocker:1995kb}
\hrefCMSnoop {} {A.~H{\"o}cker and V.~Kartvelishvili, ``{SVD} Approach to Data
  Unfolding'',} \textit{ Nucl. Instrum. Meth. A} \textbf{ 372} (1996) 469,
  \href{http://dx.doi.org/10.1016/0168-9002(95)01478-0}{\doi{10.1016/0168-9002(95)01478-0}},
\href{http://www.arXiv.org/abs/hep-ph/9509307}{\texttt{ arXiv:hep-ph/9509307}}.
%%CITATION = HEP-PH/9509307;%%.

\bibitem{Lai:2010vv}
H.-L. Lai\hrefCMSnoop {} { {et~al.}, ``New parton distributions for collider
  physics'',} \textit{ Phys. Rev. D} \textbf{ 82} (2010) 074024,
  \href{http://dx.doi.org/10.1103/PhysRevD.82.074024}{\doi{10.1103/PhysRevD.82.074024}},
\href{http://www.arXiv.org/abs/1007.2241}{\texttt{ arXiv:1007.2241}}.
%%CITATION = ARXIV:1007.2241;%%.

\bibitem{Khachatryan:2010xn}
\hrefCMSnoop {} {{ CMS} Collaboration, ``Measurements of Inclusive {W} and {Z}
  Cross Sections in {pp} Collisions at $\sqrt{s}$ = 7 {TeV}'',} \textit{ J.
  High Ener. Phys.} \textbf{ 01} (2011) 080,
  \href{http://dx.doi.org/10.1007/JHEP01(2011)080}{\doi{10.1007/JHEP01(2011)080}},
\href{http://www.arXiv.org/abs/1012.2466}{\texttt{ arXiv:1012.2466}}.
%%CITATION = ARXIV:1012.2466;%%.

\bibitem{Chatrchyan:2012ypy}
\hrefCMSnoop {} {{ CMS} Collaboration, ``Search for heavy resonances in the
  W/Z-tagged dijet mass spectrum in {pp} collisions at 7 {TeV}'',} \textit{
  Phys. Lett. B} \textbf{ 723} (2013) 280,
  \href{http://dx.doi.org/10.1016/j.physletb.2013.05.040}{\doi{10.1016/j.physletb.2013.05.040}},
\href{http://www.arXiv.org/abs/1212.1910}{\texttt{ arXiv:1212.1910}}.
%%CITATION = ARXIV:1212.1910;%%.

\bibitem{cowan1998statistical}
G.~Cowan, ``Statistical Data Analysis''.
\newblock Oxford University Press, USA, 1998.

\bibitem{efron1993introduction}
B.~Efron and R.~Tibshirani, ``An Introduction to the Bootstrap''.
\newblock Monographs on statistics and applied probabilities. Chapman \&
  Hall/CRC, 1993.

\end{thebibliography}\endgroup
